\documentclass[11pt]{article}
\usepackage{graphicx} 
\usepackage{natbib}
\usepackage[a4paper, margin=1in]{geometry}
\usepackage{natbib}
\usepackage{hyperref}
\usepackage{booktabs}
\usepackage{subcaption}
\usepackage{enumitem}
\usepackage{amsmath, amsfonts, amssymb}
\usepackage{subcaption}
\usepackage{cleveref}
\usepackage{setspace}

\title{Incorporating LLM Embeddings for Variation Across \\ the Human Genome}
\author{Hongqian Niu$^1$, Jordan G. Bryan$^2$, Jacob Williams$^3$, Hufeng Zhou$^4$, Zhun Deng$^5$, \\Haoyu Zhang$^3$, Xihao Li$^{1,6}$\footnotemark[1], and Didong Li$^{1}$\footnotemark[1]\\\\
\small Dept. of Biostatistics$^1$, Computer Science$^5$, and Genetics$^6$, Univ. of North Carolina at Chapel Hill,\\
\small School of Data Science, University of Virginia$^2$, \\ 
\small Division of Cancer Epidemiology and Genetics, National Cancer Institute$^3$, \\ \small Dept. of Biostatistics, Harvard T.H. Chan School of Public Health$^4$}
\date{}

\begin{document}
\doublespacing
\maketitle

\footnotetext[1]{Corresponding authors: \url{xihaoli@unc.edu}, \url{didongli@unc.edu}}
\footnotetext[2]{This work is supported by grants from OpenAI, Nvidia, and Hugging Face}




\begin{abstract}

Recent advances in large language model (LLM) embeddings have enabled powerful representations for biological data, but most applications to date focus on gene-level information. We present one of the first systematic frameworks to generate genetic variant-level embeddings across the entire human genome. Using curated annotations from FAVOR, ClinVar, and the GWAS Catalog, we construct functional text descriptions for 8.9 billion possible variants and generated embeddings at three scales: 1.5 million HapMap3/MEGA variants, $\sim$90 million imputed UK Biobank (UKB) variants, and $\sim$9 billion all possible variants. Embeddings were produced using general purpose models including both OpenAI’s \texttt{text-embedding-3-large} and the open-source \texttt{Qwen3-Embedding-0.6B} models. Baseline quality control experiments demonstrate high predictive accuracy for variant-level properties, validating the embeddings as structured representations of genomic variation. We further apply them to real-world embedding-augmented genetic risk predictions that demonstrate the performance of using LLM embeddings in polygenic risk score (PRS) style predictions over the UK Biobank cohort data. These resources, publicly available on Hugging Face, provide a foundation for advancing large-scale genomic discovery and precision medicine.
\end{abstract}

\section{Introduction}


   In the past few years, foundation models based on large transformer networks such as Google's BERT \citep{kenton2019bert} and OpenAI's GPT family \citep{radford2018improving} have been shown to be invaluable aids for scientific discovery in the analysis of genomic data \citep{cui2024scgpt, theodoris2023transfer, chen2025simple}. More specifically, foundation models targeted for genomic applications typically comprise of those that are trained on enormous databases of experimental data such as scGPT \citep{cui2024scgpt}, which was trained on transcriptomes from 33 million human cells from 441 different studies or the GeneFormer model \citep{theodoris2023transfer}, which was trained on 29.9 million human single-cell transcriptomes. On the other hand, foundation models based on pre-training on internet-scale databases of natural language texts may offer distinct advantages, such as potentially taking advantage of niche biological relationships which may be widely documented in scientific literature, but not necessarily be represented experimentally in large-scale genomics datasets. 

    For this reason, some recent works have used the embedding outputs of large-language models (LLMs) such as ChatGPT \citep{radford2018improving} to encode the biological information contained in text-based gene descriptions, such as those in the NCBI database \citep{schoch2020ncbi}. Notably, \cite{chen2025simple} show that these text-based gene descriptors can be input to GPT-3.5 to obtain gene embeddings that act as features/covariates for standard prediction algorithms, denoted GenePT. Furthermore, \cite{chen2025simple} showed that such embeddings can be processed at the single-cell level by taking the weighted sum of represented genes to produce relevant aggregated single-cell embeddings. For various gene-level tasks such as gene functionality class prediction, gene property prediction, and gene-gene interaction prediction, using these gene embeddings as features for a random forest algorithm was shown to have favorable predictive performance, even compared to that of specially pre-trained transformer models such as scGPT and GeneFormer, or BiolinkBERT \citep{yasunaga2022linkbert}. 
    
    Although existing LLM-based methods like GenePT have shown strong performance in genomic analysis, they primarily focus on gene-level embeddings derived from transcriptomic data. However, studying the differences in DNA sequence between individuals (genomic variation, or genetic variants) could reveal previously unknown mechanisms of human biology~\citep{igvf2024deciphering}. Genetic variants, such as single-nucleotide variants (SNVs), insertion-deletions (indels), or structural variants (SVs), play a fundamental role in uncovering the basis of genetic predispositions to diseases, and guide the development of new diagnostic tools and therapeutic targets. This highlights the need for a systematic approach to generate LLM-based embeddings for genetic variants, enabling more effective downstream scientific discovery.

    Existing works in this direction have focused on representation learning over genetic variants, including GV-Rep \citep{li2024gv}, which constructed a high quality dataset of over 7.5 million significant variants derived from ClinVar, GTEx \citep{gtex2020gtex} and experimental validation, focused primarily on curating a dataset for foundation model pre-training. Other works focus on building novel foundation models including ClinVar-BERT \citep{li2025text}, trained over plain text descriptions of 2.4 million variants for variant-level classification tasks, while others such as BMFM-DNA \citep{li2025bmfm} train over variant-aware sequence data or GPN-MSA \citep{benegas2024gpn} which incorporates evolutionary sequence data, again to predict variant-level characteristics. Relatedly, AlphaGenome \citep{avsec2026advancing} is one of the largest DNA sequence models that covers the entire human genome again for predicting variant-level characteristics. In a more similar direction to GenePT, \cite{wu2025predicting} construct natural language representations of genetic variants leveraging data from resources such as ClinVar, generated systematic embeddings using language models, and validated the utility of the derived embeddings in primarily variant-level prediction tasks of variant pathogenicity and clinical significance, using a total of 8,917 genetic variants. Finally, moving from variant-level prediction tasks to individual-level analysis, large scalable foundation models such as PRSformer \citep{dibaeinia2025prsformer} have been developed to predict phenotypic traits directly from genotype on large biobank sized studies. 

While these works establish the downstream utility of LLM and LLM embedding approaches for genetic variants, the existing literature remains bifurcated in focus and scale. Among efforts curating functional annotations, the coverage is often limited to a small fraction of the genome or restricted to variant-level classification. Conversely, while individual-level models like PRSformer achieve biobank-scale performance in phenotype prediction, they rely on representations learned directly from genotype-phenotype observations, without explicitly leveraging external functional knowledge available in comprehensive genomic resources. This leaves a gap for a comprehensive, general-purpose, whole-genome resource of functional annotation and embeddings that can serve as a foundation for diverse downstream applications.

    
    To address this gap, we have led among the first efforts in the field by curating natural language annotations for all possible 8,812,917,339 SNVs and 79,997,898  observed indels in the human genome using the FAVOR~\citep{zhou2023favor}, ClinVar~\citep{landrum2016clinvar}, and the GWAS Catalog~\citep{buniello2019nhgri, cerezo2025nhgri} databases, leading to 8,892,915,237 variants in total. From these rich resources, we generated three embedding datasets: (1) embeddings for $\sim$1.5 million SNVs in HapMap3~\citep{international2010integrating} $+$ Multi-Ethnic Genotyping Arrays (MEGA,~\citealp{bien2016strategies}) chip array; (2) embeddings for $\sim$90 million variants imputed using the Haplotype Reference Consortium~\citep{haplotype2016reference} and UK10K~\citep{uk10k2015uk10k} $+$ 1000 Genomes~\citep{10002015global} reference panels, both computed with OpenAI’s \texttt{text-embedding-3-large} model; and (3) embeddings for the full set of $\sim$9 billion possible variants using \texttt{Qwen3-Embedding-0.6B}. These resources provide one of the first systematic representations of human genomic variation at this scale, offering a unique foundation for downstream analysis.


   In addition to creating these embeddings, we develop embedding-augmented approaches for genetic risk predictions that go beyond standard polygenic risk scores (PRS). By integrating individual-level embeddings into conventional genome-wide association study (GWAS) effect-size-based PRS pipelines, we aim to improve prediction accuracy and computational efficiency. We evaluate this approach on phenotypes with clear public health relevance, including type 2 diabetes, prostate cancer, and blood lipid levels, where improved risk prediction can directly inform screening, prevention, and precision medicine strategies. Furthermore, we show that these LLM-embedding aided predictions perform comparably to and sometimes better than modern PRS methods including Clumping and Thresholding (CT) \citep{wray2007prediction, international2009commonCT}, LDPred2 \citep{prive2020ldpred2}, and Lassosum2 \citep{prive2022lassosum2}, demonstrating how these functional annotation derived embeddings can integrate semantic meaning with general pre-trained knowledge, and be broadly applied in downstream genetic analysis.

\section{Methods}

    Here we provide details on our pipeline for developing embeddings for $\sim$9 billion variants, using curated text-based annotations based on high quality data from FAVOR~\citep{zhou2023favor}, ClinVar~\citep{landrum2016clinvar}, and the GWAS Catalog~\citep{buniello2019nhgri, cerezo2025nhgri} databases. These embeddings compress rich, heterogeneous variant-level functional annotations into structured numerical representations that can be integrated into statistical genetic analyses. We also present details on our PRS-style risk prediction application by aggregating variant-level embeddings into individual-level representative embeddings. 
    

\subsection{Datasets}

    We start with the FAVOR database~\citep{zhou2023favor}, which contains functional annotation data sourced from multiple large-scale public datasets, including both public genotype databases and individual studies, for all possible 8,812,917,339 possible SNVs in the human genome (reference base with three possible single alternate alleles), as well as 79,997,898 observed indels. In total, FAVOR contains $>$160 functional annotation fields including information on variant categories, allele frequencies, integrative scores, protein functions, conservation scores, chromatin states, etc. From this, we then join data from the ClinVar \citep{landrum2016clinvar} database, which is a public archive of clinically significant variants and their associations with human disease or drug responses, based on high quality studies sourced from clinical testing laboratories, research laboratories, and other expert groups. In total, ClinVar contains high quality clinically verified annotations for over 300,000 genetic variants. Finally, we then join annotations from the GWAS Catalog \citep{buniello2019nhgri, cerezo2025nhgri}, which contains information from human GWAS and is the largest publicly available resource for GWAS results. In total, GWAS catalog contains information on over 1 million curated variant-trait associations, derived from over 7,600 different peer-reviewed publications. In summary, FAVOR provides baseline functional annotations based on published studies or in-silico machine learning model predictions for all 8.9 billion possible human variants, while ClinVar and GWAS Catalog further augment the database with high-quality peer-reviewed clinically supported and statistical associations where available.

\subsection{Annotations}

    After joining the final dataset, we then derive natural language text annotations for all 8.9 billion possible human variants similar to the NCBI database \citep{schoch2020ncbi} text summaries derived from genomic studies. Variants are identified  by chromosome, position, reference, and alternate allele, with rsIDs as applicable, under the NCBI GRCh38/UCSC hg38 build (as is FAVOR). Then, from FAVOR, we incorporate variant effect predictor categories relative to transcripts through its GENCODE category \citep{frankish2019gencode}, where it will label the gene name of the variant that has impact and if the variant is intergenic, the nearby gene name will be reported in the annotation. We also incorporate categorical MetaSVM \citep{dong2015comparison} pathogenicity predictions for disruptive missense variant, and continuous Combined Annotation Dependent Depletion (CADD) Phred-scaled scores \citep{kircher2014general} for all variants, where higher values are predicted to be more functional, pathogenic or deleterious. Furthermore, we include CAGE (Cap Analysis of Gene Expression,~\citealp{fantom2014promoter}) results, an experimental method that identifies promoter/enhancer activity, rDHS (representative DNase I Hypersensitive Sites,~\citealp{encode2012integrated}) based on the ENCODE Project, and predicted enhancer based on the GeneHancer \citep{fishilevich2017genehancer} database. Relevant clinical data and statistical associations are then included from ClinVar and GWAS Catalog, summarized in Table \ref{tab:database-fields}. 
        
        \begin{table}[ht]
        \centering
        \begin{tabular}{ll}
            \toprule
            Database & Data Fields \\
            \midrule
            FAVOR & \small Gencode Category/Info, MetaSVM, CADD-Phred, CAGE,  \\
                  & \small GeneHancer, rDHS\\
            ClinVar & \small Clinical Pathogenicity, Disease Condition, Review Confidence\\
            GWAS Catalog &  \small Disease/Trait Statistical Associations \\
            \bottomrule
            \end{tabular}
        \caption{Summary of functional annotation data used from each database to produce variant-level semantic annotations. See Appendix for full list of data fields.}
        \label{tab:database-fields}
        \end{table}

\begin{figure}[ht] 
            \centering
            \includegraphics[width=0.9\linewidth]{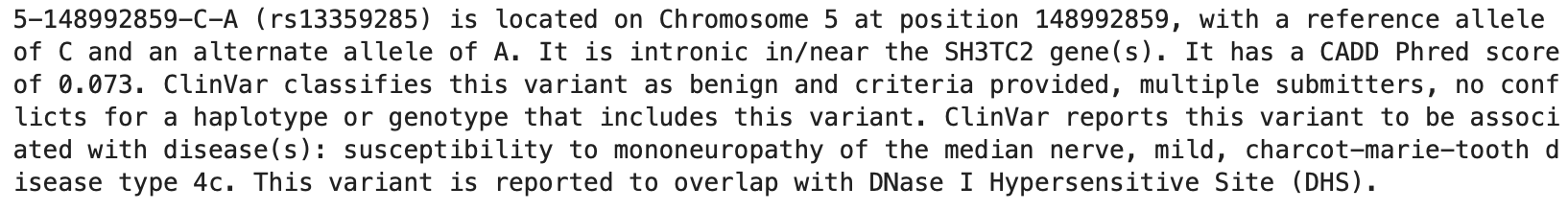}
            \caption{Sample annotation for an SNV at position 148992859 on chromosome 5 with reference allele C and alternate allele A based on the GRCh38/hg38 genome build.}
            \label{fig:anno-example}
            \end{figure}

    Figure \ref{fig:anno-example} presents a sample annotation for a particular variant, with more examples and descriptive statistics provided in the next section. 

\subsection{Variant-Level LLM Embeddings}

    From the variant-level functional annotations, we then derive LLM embeddings from both open-source embedding models such as \texttt{Qwen3-Embedding-0.6B}, as well as \texttt{text-embedding-3-large} available through the OpenAI API (we note that OpenAI does not currently support embedding models beyond GPT-3.5). Balancing scientific significance and accessibility for practical use, from this we derive the following sets of variant-level embeddings at three different scales as described in the introduction:
       \begin{enumerate}[itemsep=0pt]
          \item \textbf{HapMap3 \& MEGA} ($\sim$1.5~million variants)
          \item \textbf{UK Biobank Imputed} ($\sim$90~million variants)
          \item \textbf{All FAVOR Variants} ($\sim$9~billion variants)
          \end{enumerate}


    Although we may ideally also generate embeddings for the full 9 billion variants dataset using \texttt{text-embedding-3-large} for comparison, practically the vast majority of possible variants are understudied, with limited information currently present even in the largest of public repositories. To this extent, both the 1.5 million SNV and 90 million variant datasets consist of primarily well-studied variants such as those that are given rsIDs, are more rich in their functional annotations, and are directly applicable to analysis on the individual-level genotype data (e.g. from the UKB cohort). We see in downstream analysis on PRS-style prediction tasks in Table \ref{tab:UKB-pred-continous-r2} and Table \ref{tab:UKB-pred-binary-auc} that the significantly smaller and less computationally intensive \texttt{Qwen3-Embedding-0.6B} actually performs comparably, meanwhile we estimate an API cost of approximately $\sim$\$100,000, to embed the entire FAVOR variants dataset. As such, we rely on using solely the open-source models for the full 8.9 billion dataset, with more details on computational costs for the analysis presented in the Appendix. 

\subsection{Individual-Level Embeddings} \label{sect:indiv-level-emb}

    As part of the pipeline for phenotypic studies using the UKB cohort, we generate individual-level embeddings for each individual from the study by taking an average over the variant-level embeddings, weighted by each individual's genotype dosage (originally 0, 1, or 2) based on the UKB genotype reference alleles. For a given embeddings matrix $\mathrm{E} \in \mathbb{R}^{p\times d}$, where $p$ is the number of genetic variants of interest and $d$ is the dimension of the LLM embeddings, and an individual genotype matrix $\mathrm{G}\in \mathbb{R}^{n\times p}$ where $n$ is the total number of individuals, then the final individual level embedding is simply a product of the two matrices $\mathrm{L} = \mathrm{G}\mathrm{E} \in \mathbb{R}^{n\times d}$. 

    Furthermore, we also explore the use of GWAS-weighted individual level embeddings. Rather than weighting the embedding solely by the genotype dosage, we also multiply each genetic variant's dosage by the signed marginal effect size from prior GWAS study, in a manner similar to traditional PRS methods discussed below. Hence for each set of LLM-derived variant embeddings, we generate two sets of individual-level embeddings: (1) using the genotype dosage directly as weights, and (2) using genotype dosage further multiplied by GWAS estimated effect sizes as weights, for use in different tasks.
    
    Finally, we note here that the number of genetic variants $p$ in a study is often far greater than the dimension $d$ of the LLM embeddings. In the UKB cohort, individual genotypes were imputed to a total of over 90 million variants (of which HapMap3 \& MEGA variants are a subset), whereas the LLM embeddings we used ranged from 1,024 to 3,072 dimensions, depending on the embedding model. 

\subsection{PRS-Style Phenotype Prediction}

    The typical task of PRS is to quantify the effect of genome-wide genetic variants on a quantitative or disease phenotype of interest such as blood lipid levels or cancer status, capturing important polygenic contributions across a large number (e.g. thousands) of variants. However, direct modeling of genetic risk using the genotype dosage is challenging due to the significant number of variants required to be modeled in order to capture the impact of many variants of small effects contributing to phenotypic disease risk. Traditional techniques for PRS predictions hence take advantage of GWAS summary statistics to use marginal effect sizes and associated $p$-values to either directly filter variants based on statistical significance or are treated as prior information in penalized regression models or Bayesian modeling frameworks. In this work, we compare our PRS-style prediction to three modern PRS algorithms, namely Clumping and Thresholding (CT) \citep{wray2007prediction, international2009commonCT}, LDpred2 \citep{prive2020ldpred2}, and Lassosum2 \citep{prive2022lassosum2}. 
    
    Briefly, CT selects representative variants based on linkage disequilibrium (LD) blocks to reduce correlation effects, and then thresholds variants based on GWAS statistical significance. The result is a score based on a filtered set of variants using the original estimated marginal GWAS effect sizes. In contrast, LDpred2 is a modern Bayesian approach which models LD explicitly and computes posterior mean effect sizes conditional on both the GWAS summary statistics and LD structure. Finally, Lassosum2 is a penalized regression model which uses LD and GWAS summary statistics to produce regularized, joint variant effect estimates over the set of all variants, under Lasso shrinkage. While CT is oftentimes the quickest and most straightforward approach, LDpred2 and Lassosum2 notably re-estimate the effect sizes and often improve on the performance of CT, but may take longer for computation. 
    
    Our pipeline similarly takes advantage of GWAS summary statistics, first using the marginal associations to filter for the variants with the highest statistical significance. In practice, the specific $p$-value threshold chosen does not need to be strict as in GWAS studies, and can be treated as a hyperparameter as in traditional PRS pipelines. LD can also be incorporated in this step to reduce correlation in the filtered set of variants. After GWAS filtering, we then generate the genotype-weighted individual-level embeddings matrix $\mathrm{L}\in \mathbb{R}^{n \times d}$, which incorporates the LLM-derived embeddings capturing background knowledge of each variant from the LLM pre-training in addition to the annotations. Hence, regardless of the number of filtered variants, the final dimension of the data depends solely on the chosen LLM. Finally, from the individual-level embeddings, we can fit any machine learning model to perform regression or classification for individual phenotype prediction, treating the embeddings as features. We note here that the LLM embeddings are pre-computed and saved as a dataset at the variant level as described previously, released publicly with link in the Appendix, and can be easily re-applied for any general purpose downstream analysis where the individual genotype is known.

\section{The Variant-Foundation-Embeddings Dataset}

    Here we provide descriptive statistics and simple experiments to demonstrate baseline performance for the 1.5 million SNV annotations and the accompanying embeddings datasets. 

\subsection{Annotations}
    As described previously, we derive semantic-based annotations for the 1.5 million SNVs relevant to the UKB study based on FAVOR, ClinVar and GWAS Catalog matched on rsID, chromosome number, position, and specific reference/alternate alleles, accounting for potential reference/alternative allele flips between the UKB genotypes and FAVOR. In Table \ref{tab:UKB-anno-token-summary}, it can be seen that the annotation lengths range from 64 to 356, with a mean of 89 tokens from the \texttt{cl100k\_base} tokenizer as used by OpenAI's family of embedding models.

    \begin{table}[ht]
        \centering
        \begin{tabular}{c c c c }
            \toprule
            Min. & Max. & Mean & Std. Dev.  \\
            \midrule
            64 & 356 & 89 & 23 \\
            \bottomrule
            \end{tabular}
        \caption{Token counts for text-based annotations for the 1.5 million SNVs relevant to the UKB dataset based on rsID, reference, and alternate allele matching to FAVOR.}
        \label{tab:UKB-anno-token-summary}
        \end{table}

\begin{figure}[ht] 
            \centering
            \includegraphics[width=0.5\linewidth]{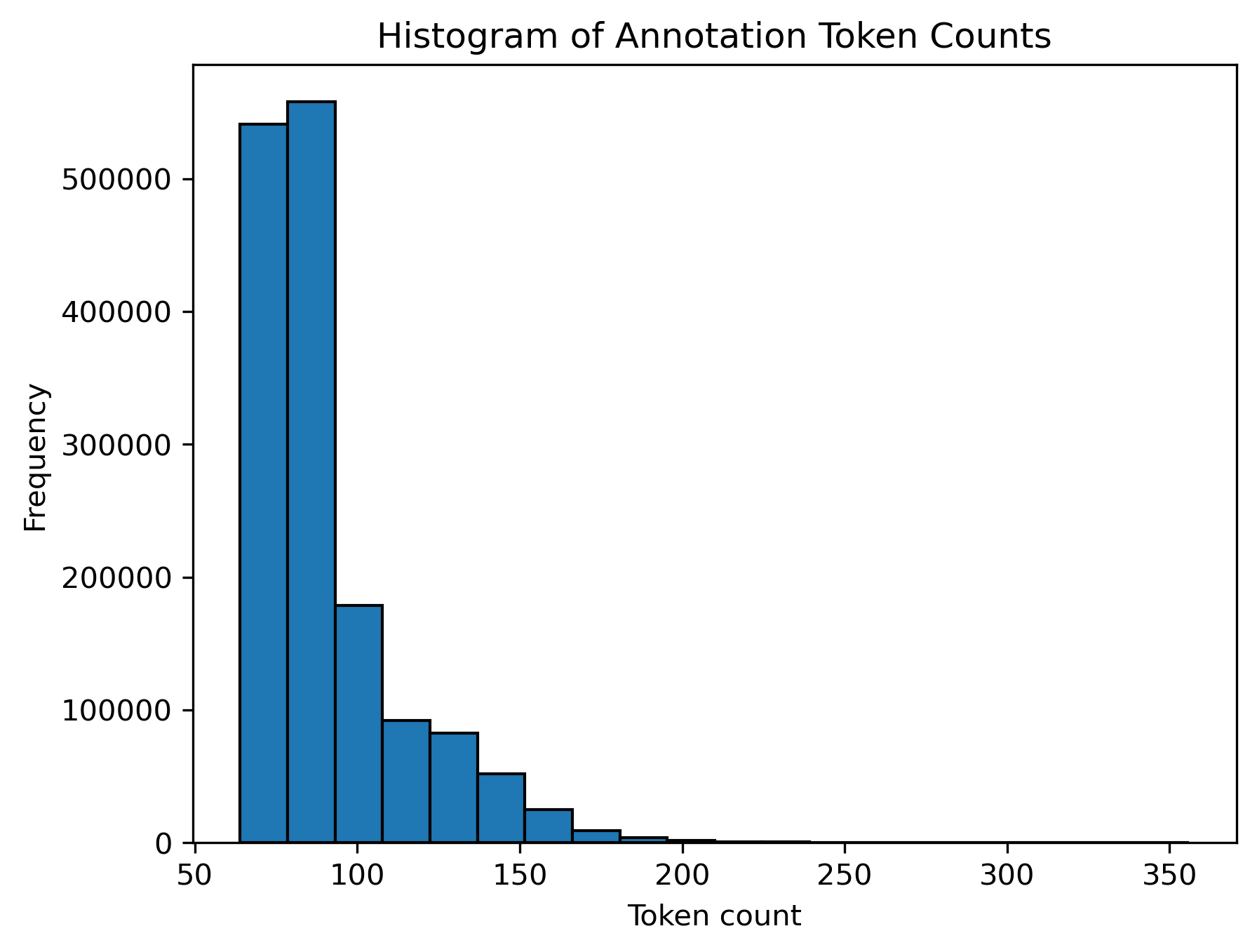}
            \caption{Histogram of token counts for derived annotations for the UKB 1.5M SNV list.}
            \label{fig:anno-token-hist-UKB}
            \end{figure}

\begin{figure}[ht!] 
            \centering
            \includegraphics[width=0.8\linewidth]{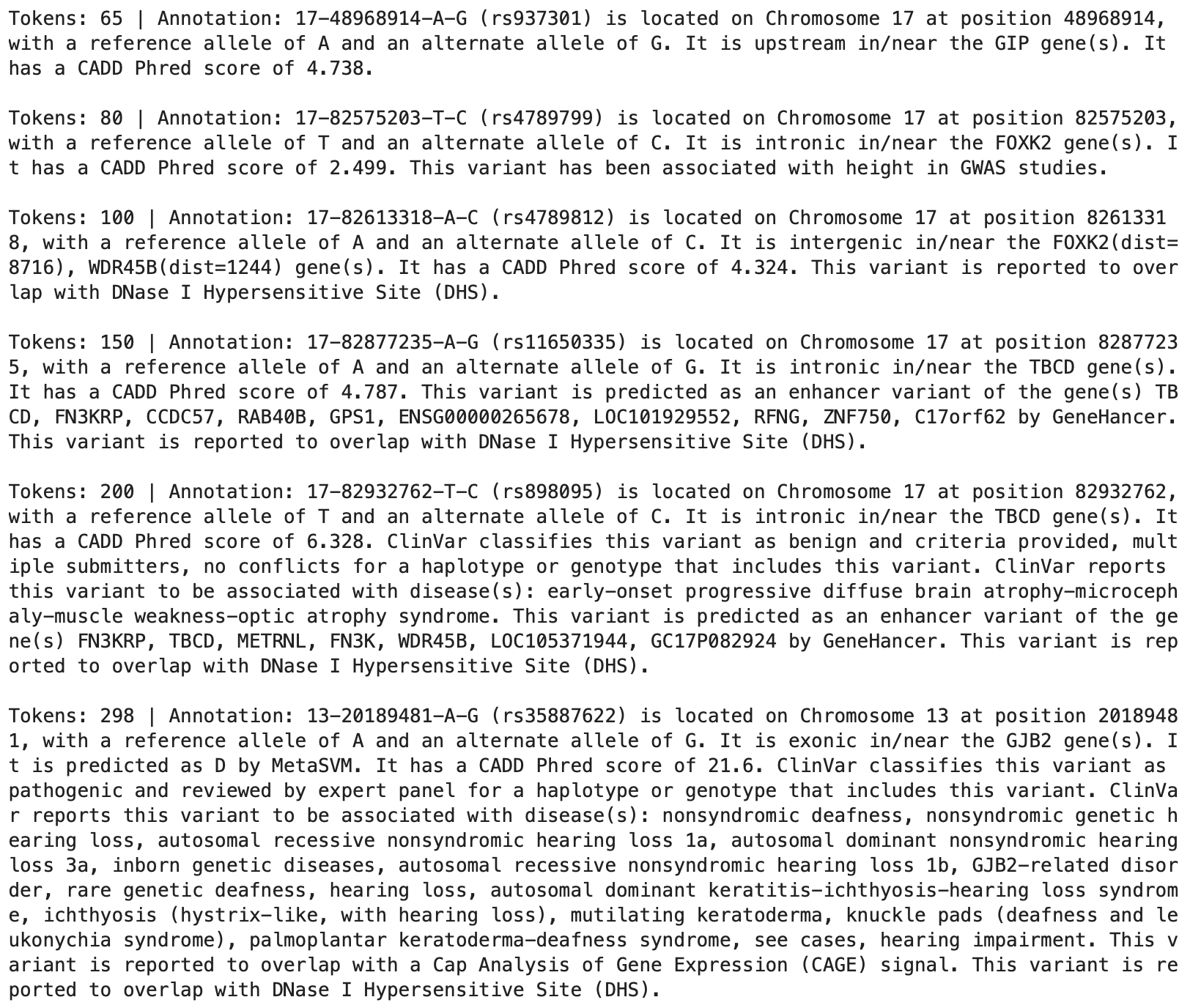}
            \caption{Examples of variant annotations at different lengths and supporting information.}
            \label{fig:anno-examples-UKB}
            \end{figure}

    This relatively large range emphasizes the varying degree of current knowledge across the vast majority of SNVs. In Figure \ref{fig:anno-token-hist-UKB}, we present the histogram of token counts within this dataset, showing the vast majority of SNVs lie below 100 tokens in the curated annotations. Of these curated annotations, the vast majority rely on more basic functional annotations from FAVOR such as relation to nearby genes or machine learning derived annotation scores. However, there are many SNVs that also contain rich clinical or statistical associations as illustrated in Figure \ref{fig:anno-examples-UKB}. 

    In Figure \ref{fig:anno-examples-UKB}, we present examples of curated annotations at varying lengths, with varying degrees of supporting information. It can be seen that in particular, ClinVar can add rich clinical information on both disease associations as well as the strength of the reporting where available. 
    
\subsection{Embeddings}\label{sec:exp}

\begin{figure}[ht!]
        \centering
        \begin{subfigure}{0.45\linewidth}
            \centering
            \includegraphics[width=\linewidth]{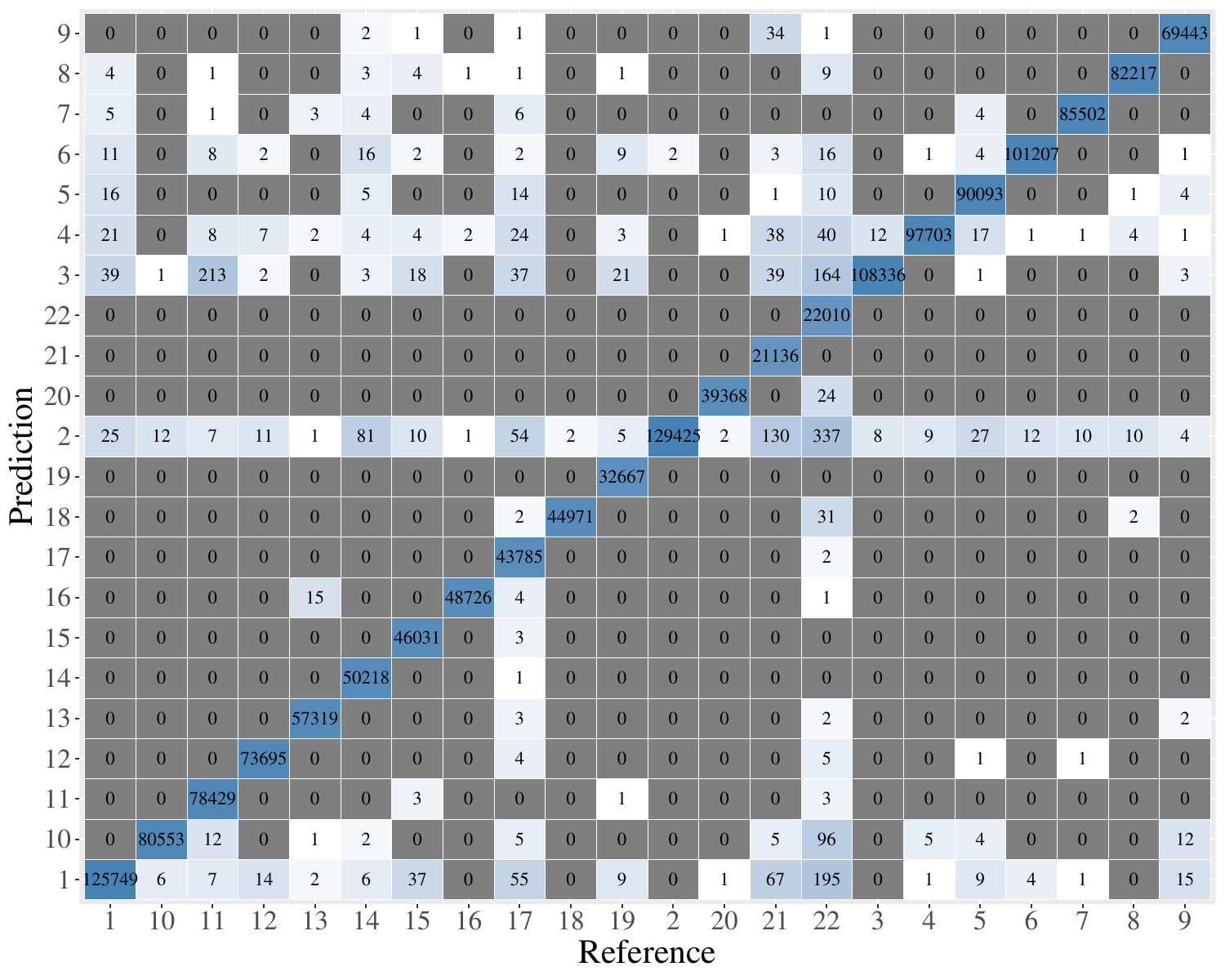}
            \caption{OpenAI Text-Embedding-3-Large}
            \label{fig:cfm-gpt3-a}
            \end{subfigure}
        \begin{subfigure}{0.45\linewidth}
            \centering
            \includegraphics[width=\linewidth]{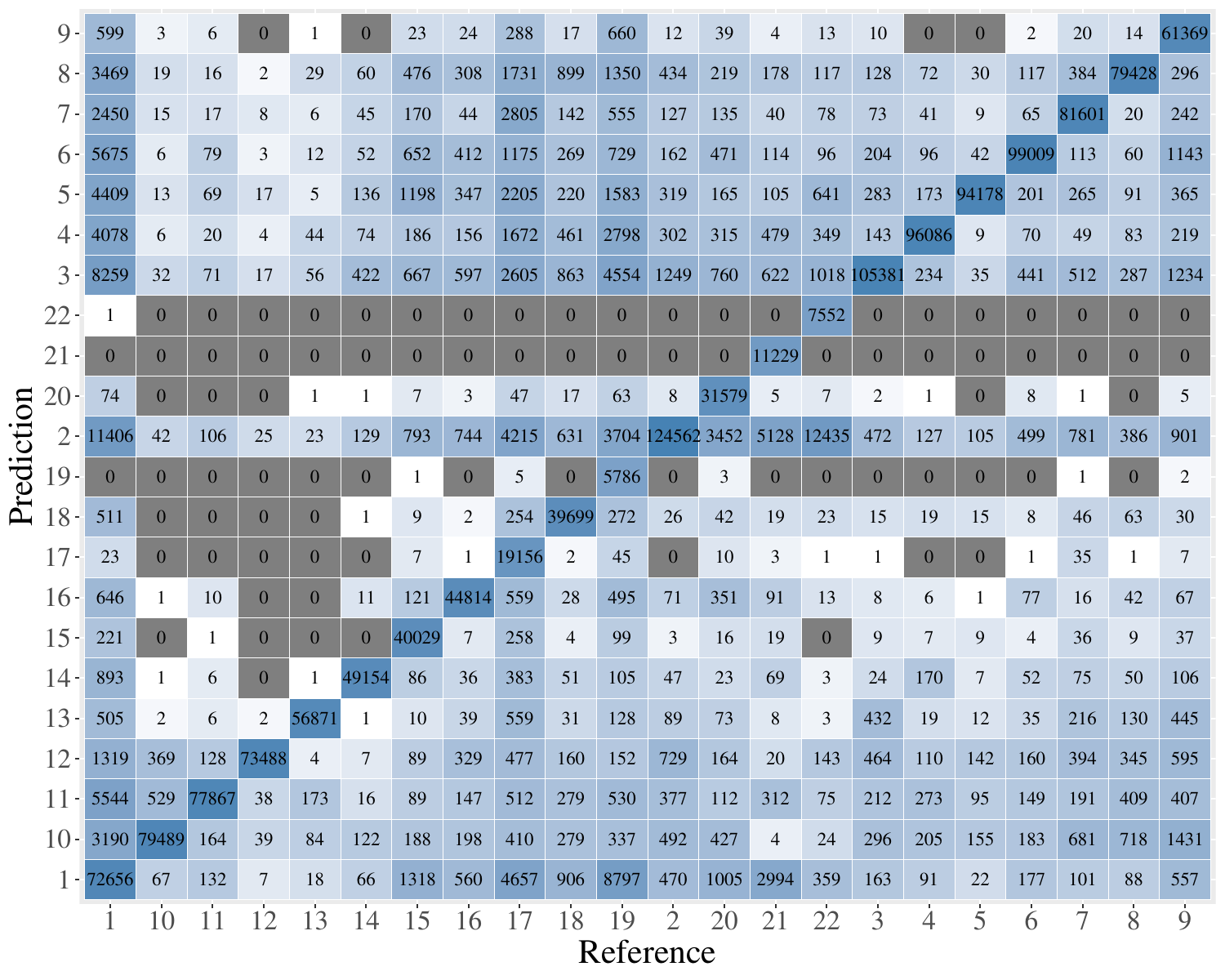}
            \caption{Qwen3-Embedding-0.6B}
            \label{fig:cfm-qwen06b-b}
            \end{subfigure}
        \caption{Chromosome number prediction task using variant-level embeddings from a) OpenAI \texttt{text-embedding-3-large} with prediction accuracy of greater than 99\%, and b) \texttt{Qwen3-Embedding-0.6B} with prediction accuracy of 88\%.}
        \label{fig:chrom-both}
        \end{figure}

    Next we consider the set of embeddings derived from these semantic-based variant annotations, using the OpenAI \texttt{text-embedding-3-large} and \texttt{Qwen3-Embedding-0.6B} embedding models. In each case, we keep the full native embedding vectors where available, which are 3,072-dimensional for OpenAI and 1,024-dimensional for Qwen3-0.6B.  Hence, the size of the final embeddings dataset is 1.5 million by 3,072 or 1,024 embedding dimensions. In Figure \ref{fig:chrom-both}, we build random forest classifiers to predict the chromosome number using a random sample of 10,000 variants' embeddings as the training data, for each set of embeddings (OpenAI and Qwen), and report predictions on all remaining variants in the dataset. It can be seen that accuracy is near perfect across all 22 tested chromosomes for the OpenAI embedding model, and also highly accurate at 88\% for the Qwen3-0.6B model. 

\begin{figure}[ht!]
        \centering
        \begin{subfigure}{0.4\linewidth}
            \centering
            \includegraphics[width=\linewidth]{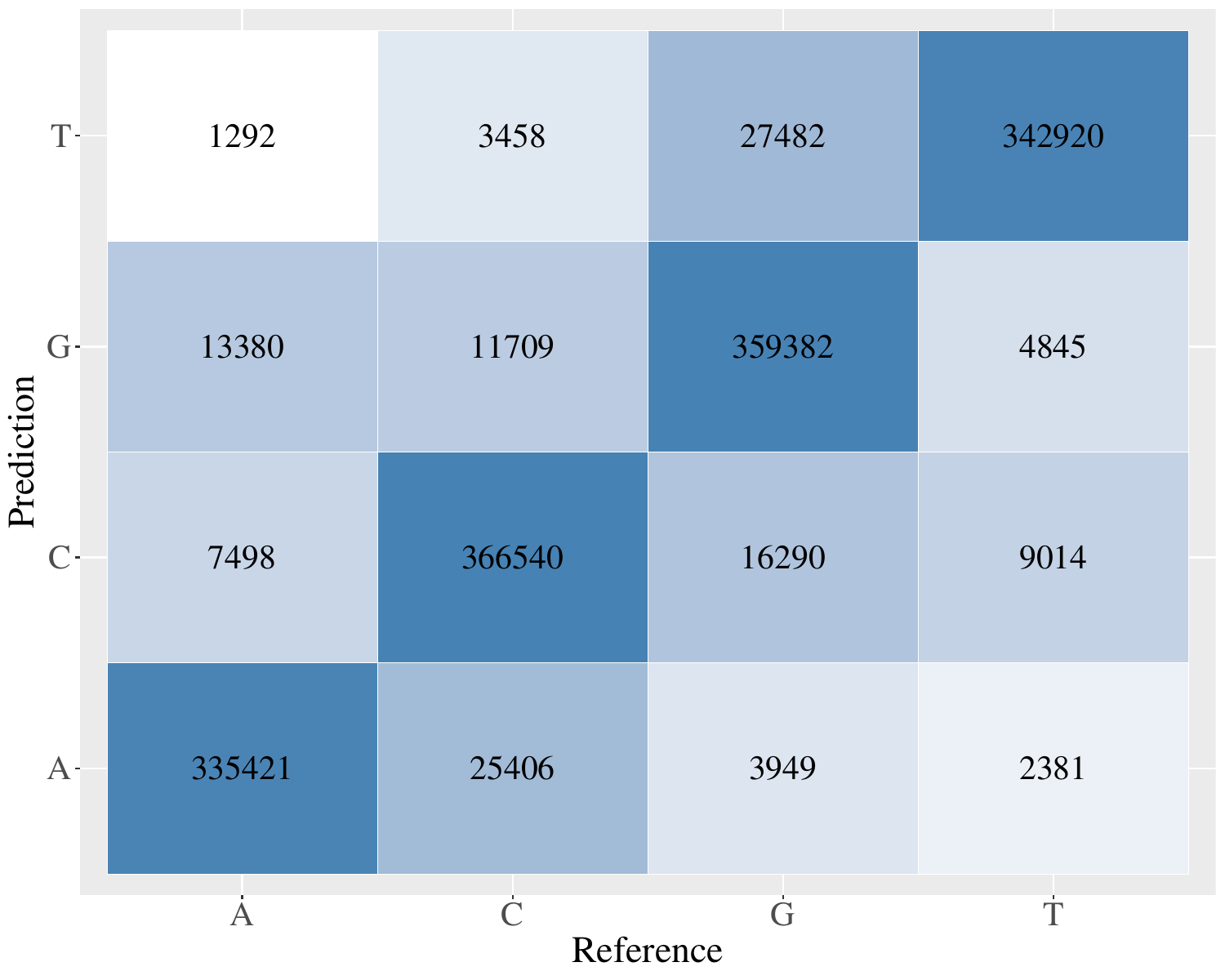}
            \caption{OpenAI Text-Embedding-3-Large}
            \label{fig:ref-gpt3-a}
            \end{subfigure}
        \begin{subfigure}{0.4\linewidth}
            \centering
            \includegraphics[width=\linewidth]{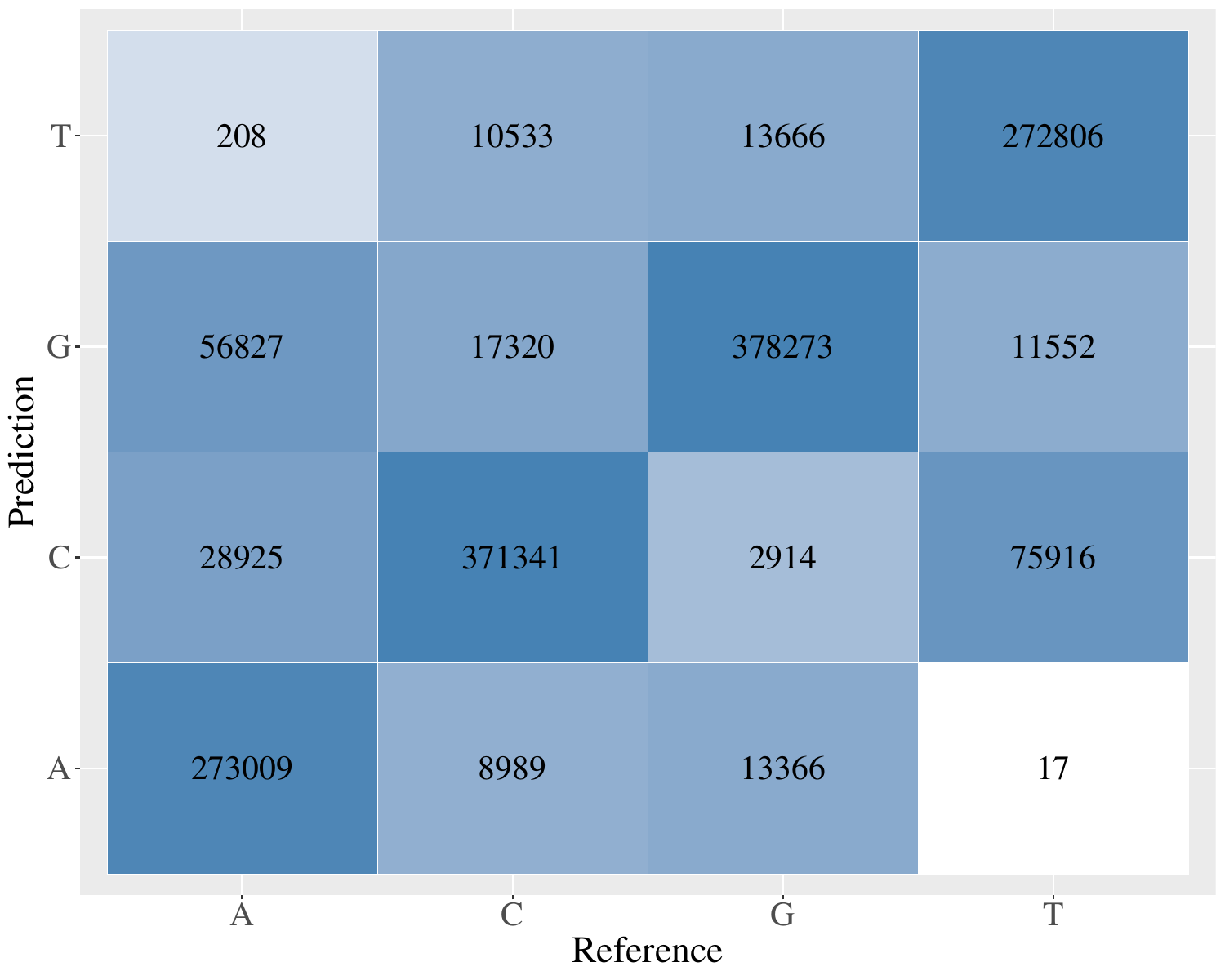}
            \caption{Qwen3-Embedding-0.6B}
            \label{fig:ref-qwen06b-b}
            \end{subfigure}
        \caption{Reference allele prediction with a) OpenAI \texttt{text-embedding-3-large} with prediction accuracy of 92\% and b) \texttt{Qwen3-Embedding-0.6B} with prediction accuracy of 86\%.}
        \label{fig:ref-both}
        \end{figure}

    Again in Figure \ref{fig:ref-both}, we see that the embeddings are also predictive of the reference allele with overall 92\% and 86\% accuracy for the OpenAI and Qwen3-0.6B models respectively, while  trained on only a random subsample of 10,000 out of 1.5 million SNVs.

     While these prediction tasks are designed to be straightforward and test only data that were explicitly present in the annotations, they serve as baseline quality checks to ensure the embeddings are correctly capturing at least what is in the annotations. These SNV-level embeddings then form the basis for downstream genetic studies such as PRS-style prediction of complex traits or diseases. 

\subsection{UK Biobank Cohort}
     The UKB is a large cohort study that recruited over 500,000 individuals aged 40-69 between 2006 and 2010, collecting extensive phenotype information associated with large-scale genotyping  \citep{bycroft2018uk}. For this study we use a subset of the cohort with 140,488 unrelated individuals for which both phenotype and genotype data are available, accounting for of 1,546,684 SNVs and a range of continuous and binary traits such as height, high-density lipoprotein cholesterol levels, cancer status, etc. 

\section{Results}

    \subsection{GWAS Ancestry Adjustment} \label{sec:GWAS-ancestry-adjustment}
    Biobanks generally aggregate data over a large, diverse sample of individuals, hence it is typical and often necessary to adjust for ancestry when performing statistical analyses such as GWAS. In this study, we consider both binary and continuous GWAS for 11 traits, including five binary traits (asthma, breast cancer, coronary artery disease (CAD), prostate cancer, and type 2 diabetes (T2D) status) and six continuous traits (height, total cholesterol (TC), high-density lipoprotein cholesterol (HDL), low-density lipoprotein cholesterol (LDL), log triglycerides (logTG), and body mass index (BMI)), with rich genetic data in the UK Biobank cohort. For binary traits, in the simplest case we may seek to model the marginal  associations for each genetic variant:
        \begin{align*}
            \mathrm{logit}\{ \mathrm{P}(y_i=1|\mathbf{x}_i,\mathrm{G}_{ij})\} &= \mathbf{x}_i^\top\mathbf{\alpha}+ \mathrm{G}_{ij}\beta_j \\
                                                 & = (\mathrm{age}_i, \mathrm{age}^2_i, \mathrm{sex}_i, \mathrm{PC}_{1_i}, \ldots \mathrm{PC}_{10_i})^\top\mathbf{\alpha} + \mathrm{G}_{ij}\beta_j
            \end{align*}
    where $y_i$ is the phenotype for individual $i$, $\mathrm{G}_{ij} \in \{0, 1, 2\}$ is the genotype dosage for individual $i$ at variant $j$, with associated regression coefficients $\beta_j$, and $\mathbf{x}_i$ is a vector of covariates consisting of both individual characteristics as well as ancestry principal components (PCs) derived from the whole cohort's genotype matrix, with $\mathbf{\alpha}$ the vector of associated regression coefficients. More specifically, in this study, the covariates used are age, $\mathrm{age}^2$, sex, and the top 10 ancestry PCs released by the UK Biobank for each individual $i$. In practice, we also use leave-one-chromosome-out adjustment for polygenic modeling, but here we draw attention to the role of ancestry PCs in these marginal tests of association in this simplified model. 
    
    The ancestry adjustment is typically achieved by performing principal component analysis (PCA), a linear dimension reduction method, on the raw genotype dosage matrix of individuals and variants of interest and adjusting subsequent models by the top principal components, which are often highly correlated with the ancestry substructure of the data. However, in practical applications, the genotype dosage matrix $G \in \mathbb{R}^{n\times p}$ is a computationally challenging target even for linear dimension reduction methods. Standard PCA via singular value decomposition has a runtime of $O(np\cdot\min\{n,p\})$, which in the GWAS scenario typically resolves to $O(n^2p)$ since the number of variants $p$ is frequently far larger than the cohort size $n$. In the UK Biobank, we use the cohort of $n=140,488$ individuals, while the number of variants $p$ is approximately 1.5 million. This makes direct computation challenging even for truncated/randomized PCA algorithms which can run in $O(npk)$ where $k$ is the number of top principal components. Due to the computational challenge of computing principal components, especially when the set of observed variants can be even greater than the sample size in the cohort, biobanks will typically pre-compute the ancestry PCs over the entire cohort for once and release them centrally along with the full dataset.

    This approach, however, may pose critical data leakage issues with downstream analysis. In the subsequent PRS studies, it is necessary to first split up the data into a train/tune/validation split. Nevertheless, when adjusting models for ancestry using the pre-computed PCs, there is a potential for data leakage as the ancestry PCs are often computed against the entire cohort using the genotype dosages for each individual, which is a predictor for PRS models themselves. Hence, there is a direct need for computationally efficient alternatives to compute ancestry PCs that capture ancestry substructure and can be efficiently re-computed to fit dynamic training environments of any set of individuals in the cohort. 

    \begin{figure}[ht!] 
            \centering
            \includegraphics[width=0.99\linewidth]{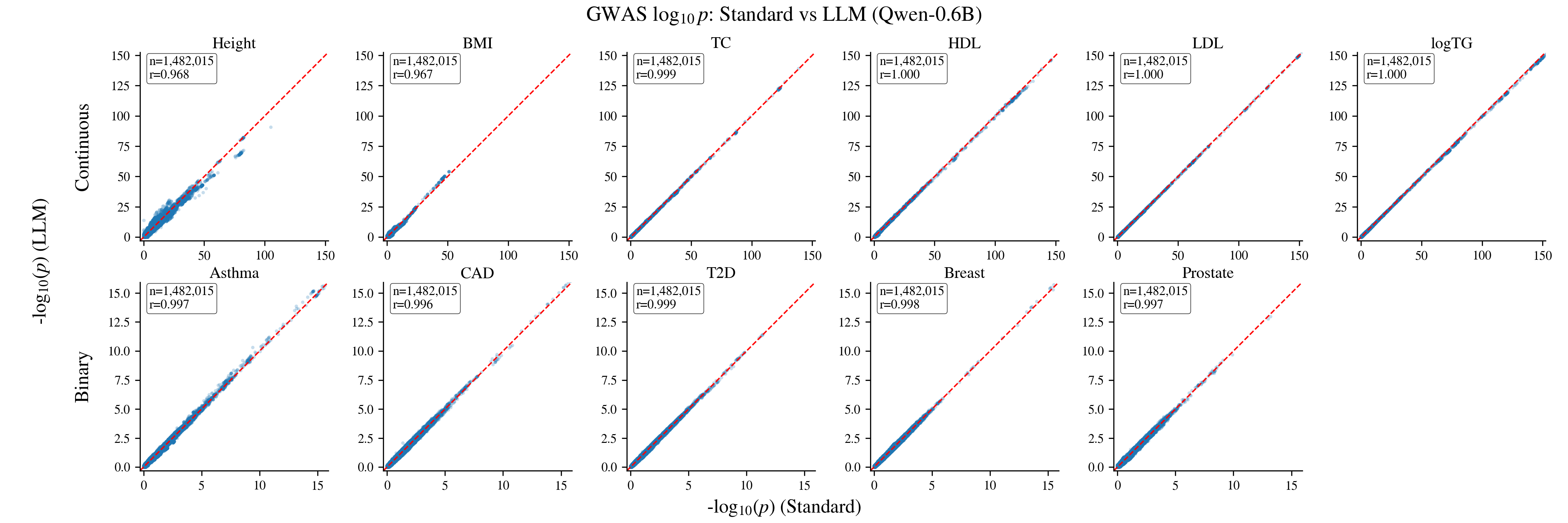}
            \caption{$-\log_{10}(p)$ comparison between standard ancestry adjustment PCs released by UK Biobank compared to $-\log_{10}(p)$ based on Qwen-0.6B embeddings over all SNPs.}
            \label{fig:GWAS-2row-qwen06}
            \end{figure}

    \begin{figure}[ht!] 
        \centering
        \includegraphics[width=0.99\linewidth]{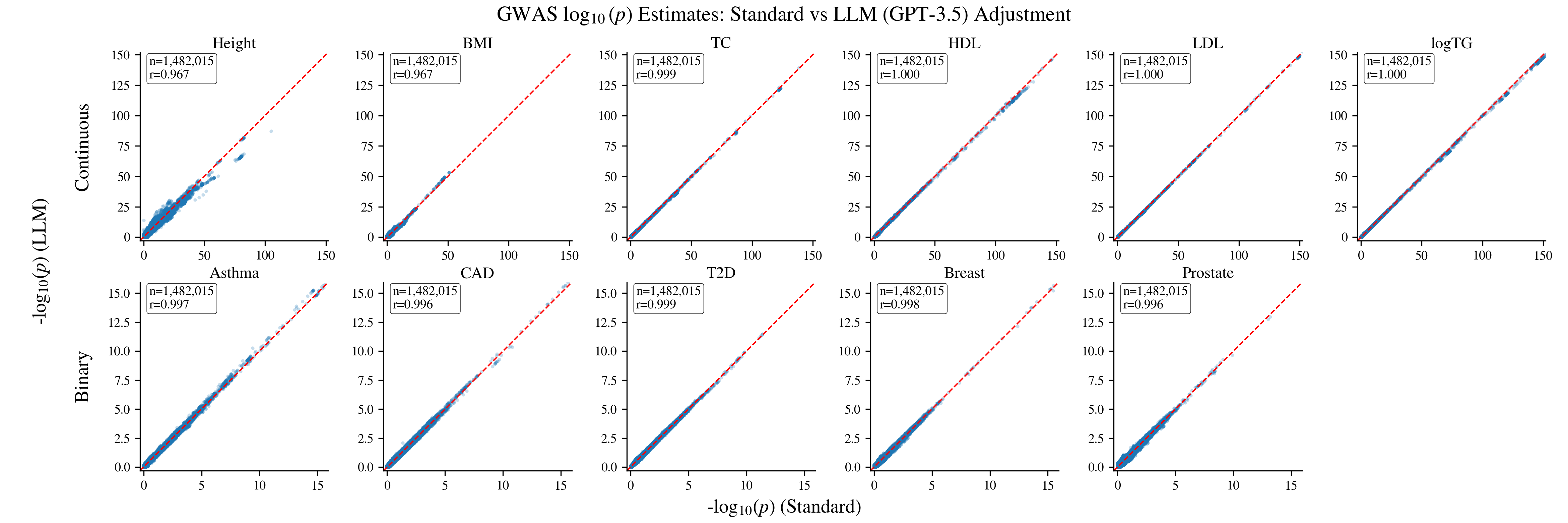}
        \caption{$-\log_{10}p$ comparison between standard ancestry adjustment PCs released by UK Biobank compared to $-\log_{10}p$ based on GPT-3.5 embeddings over all SNPs.}
        \label{fig:GWAS-2row-gpt3}
        \end{figure}

    As described earlier, our LLM based variant-level embeddings can be leveraged to produce individual level embeddings $\mathrm{L}\in \mathbb{R}^{n\times d}$, where $d$ is the dimension of the LLM generated embeddings, typically ranging between 1024 and 3072 as is in our case. Notably, $d$ is a fixed property inherent to the architecture of a specific LLM model, hence it does not scale with the number of observed variants $p$. In most real world scenarios, $d$ is often hundreds of times smaller than $p$,  which directly allows for a far more efficient way to compute analogous ancestry PCs. This effectively drops the dependence on $p$ when performing PCA as the dimension of the LLM embedding does not scale with the number of variants and can be used for any set of variants by parallelizable computation of the individual-level representation. We note that in this case we only use the genotype dosage to weight the variant-level embeddings and do not incorporate previous GWAS information into the individual-level representations. Finally, we verify that our ancestry PCs computed over the individual-level LLM embeddings, which runs on the order of seconds in this case, gives near perfect alignment with GWAS-estimated $p$-values across all 1.5 million observed SNVs as seen in \Cref{fig:GWAS-2row-qwen06} and \Cref{fig:GWAS-2row-gpt3}.

    \subsection{PRS-Style Predictions}

    Next we present the results of our PRS-style prediction pipeline aided by two separate popular models of LLM embeddings applied to the aforementioned six continuous traits and five binary traits from the UK Biobank cohort. 

    Prior to modeling, we split the cohort into a 70/15/15 train ($n=98,343$), validation ($n=21,073$), and held-out test ($n=21,072$) split. All algorithms are trained and tuned on the exact same train/tuning/validation splits, with final held-out test performance assessed at the end. The train test consists solely of European ancestry individuals, while the tuning and validation sets are mixed ancestry, although we report final validation metrics for all methods over the European subset only due to low sample size in the other ancestry groups in the UK Biobank cohort. 
    
    For each set of SNV embeddings, we compute individual level-embeddings for the UKB cohort in two ways. First, taking a weighted sum of the SNV embeddings by the individual's raw genotype dosage vector consisting of  $\{0, 1, 2\}$ entries, multiplied by the estimated effect size from GWAS studies over only the training data, as also used by the standard PRS methods. Second, we also try standardizing the genotype dosages per SNV, then multiplied by the estimated effect size from GWAS studies. The number of SNVs used is tuned via a $p$-value cutoff based on the GWAS estimated effects. Then, using the individual-level embeddings, we fit linear and logistic regressions with ridge penalty, performing hyperparameter tuning over the penalization parameter and the aforementioned grid of $p$-value cutoffs over the tuning set.
    
    Finally, for each phenotype, we compare prediction performance to CT, LDpred2, and Lassosum2 using residual $R^2$ and covariate-adjusted AUC where applicable. We note that each of these methods performs similar hyperparameter tuning over the GWAS $p$-value cutoffs, as well as uses the effect size estimates from the training split GWAS analysis. 

    \subsubsection{UK Biobank Continuous Traits}\label{sec:PRS-UKB-continuous}

    For the continuous traits: height, TC, HDL, LDL, logTG, and BMI, all prediction models are adjusted for age, $\text{age}^2$, sex, and 10 ancestry PCs as is typical in GWAS and PRS studies. For reference, we briefly review the CT algorithm. In CT, we first define a set of significance thresholds $T_k$, which determine the set of variants $\mathcal{S} (T_k) = \{j \in \mathcal{S} : p_j \leq T_k\}$ based on estimated $p$-values $p_j$ from the GWAS-derived summary statistics over the training split for SNV $j$. Here we define the set of $T_k$ to range from $\{ 5\times10^{0}, 5\times10^{-1}, \ldots, 5\times10^{-8}\}$. In practice, $\mathcal{S}$ also incorporates LD pruning to remove highly correlated variants. Then, the PRS for each individual $i$ at $p$-value cutoff $T_k$ is formed:
        \begin{align*}
            \mathrm{PRS}_i^{T_k} = \sum_{j\in S(T_k)} \widehat{\beta}_j \mathrm{G}_{ij}
            \end{align*}
    where $\mathrm{G}_{ij}$ is the genotype dosage of individual $i$ at variant $j$, and $\widehat{\beta}_j$ is the GWAS-derived effect size for variant $j$ (no model re-fitting). Assessment on the genetic trait prediction task is evaluated by first fitting the null model: 
        \begin{align*}
            Y_i & = \alpha_0 + X_i^\top\alpha+\epsilon_i \\ 
                & = \alpha_0 + (\mathrm{age}_i, \mathrm{age}^2_i, \mathrm{sex}_i, \mathrm{PC}_{1,i}, \ldots \mathrm{PC}_{10,i})^\top\alpha + \epsilon_i
            \end{align*}
    where $\alpha$ are the regression coefficients of the null model and $\epsilon_i$ the noise term. Finally, the PRS are regressed against the residuals $\widehat{r}_i = Y_i-(\widehat{\alpha}_0 + X_i^\top\widehat{\alpha})$ of the null model to evaluate the incremental $R^2$ due to PRS at $p$-value threshold $T_k$:
        \begin{align*}
            \widehat{r}_i = \gamma_0+\mathbf{\gamma}_k\mathrm{PRS}_i^{T_k}+\eta_i
            \end{align*}
    where $\mathbf{\gamma}_k$ are the regression coefficients for the PRS regression at threshold $T_k$ and $\eta_i$ the noise term. The optimal threshold is tuned using a typical training/tuning split and evaluated on a held-out validation set. 
        
    Paralleling the CT pipeline, we generate LLM-derived embeddings matrices $\mathrm{L}_{ij}^{T_k} \in \mathbb{R}^{n\times d}$ for $n$ individuals and LLM embedding dimension $d$, where the set of variants are thresholded at $p$-value thresholds $T_k$ based on GWAS summary statistics, as described in \Cref{sect:indiv-level-emb}. The LLM-aided predictions are then made using a ridge regression treating the embeddings as features, regressed directly against the residuals of the same null model adjusting for the same set of baseline covariates. More specifically, we fit directly:
        \begin{align*}
                \widehat{r}_i = \theta_k\mathrm{L}_i^{T_k}+\eta_i
                \end{align*}
    where $\widehat{r}_i$ is defined previously, $\mathrm{L}_i^{T_k} \in \mathbb{R}^d$ is the LLM-derived embedding vector for individual $i$ at threshold $T_k$, $\theta_k$ the regression parameters for the embeddings, and $\eta_i$ the noise term. Similarly, the optimal threshold is tuned using the same training/tuning split and evaluated on the same held-out validation set. Finally, due to the low sample size of other ancestry groups in the UKB cohort, we report evaluation metrics over individuals of European ancestry in the held-out validation set for all methods.

    \begin{table}[ht!]
        \centering
        \label{tab:r2_methods}
        \begin{tabular}{l c c c c c c c}
            \toprule
            Method                & Height & TC      &  HDL   & LDL    & logTG  & BMI \\
            \midrule
            CT                    & 0.212  & 0.069   & 0.098  & 0.080  & 0.061  & 0.042    \\                     
            LDpred2               & 0.292  & 0.087   & 0.156  & 0.102  & 0.105  & 0.072     \\
            Lassosum2             & 0.278  & 0.089   & 0.144  & 0.099  & 0.095  & 0.067     \\
             \hline
            Qwen-3 + $\beta_{\text{GWAS}}$        & 0.233  & 0.078   & 0.122  & 0.094  & 0.074  & 0.050 \\
            GPT-3.5 + $\beta_{\text{GWAS}} $      & 0.243  & 0.081   & 0.123  & 0.097  & 0.080  & 0.026 \\
            Qwen-3 + $\beta_{\text{GWAS-std}} $   & 0.231  & 0.080   & 0.124  & 0.095  & 0.076  & 0.049        \\
            GPT-3.5 + $\beta_{\text{GWAS-std}}$    & 0.234  & 0.081   & 0.126  & 0.104  & 0.079  & 0.026 \\
            \bottomrule
        \end{tabular}
        \caption{Predictive performance (residual $R^2$) of different PRS methods in the held-out test set over individuals of European ancestry, adjusting for baseline covariates compared to LLM-derived embeddings based methods. LLM-derived embeddings include both those using non-standardized and standardized genotype dosages for comparison. The specific models used are \texttt{Qwen3-Embedding-0.6B} and \texttt{text-embedding-3-large}.}
        \label{tab:UKB-pred-continous-r2}
    \end{table}

\begin{figure}[ht!] 
            \centering
            \includegraphics[width=0.80\linewidth]{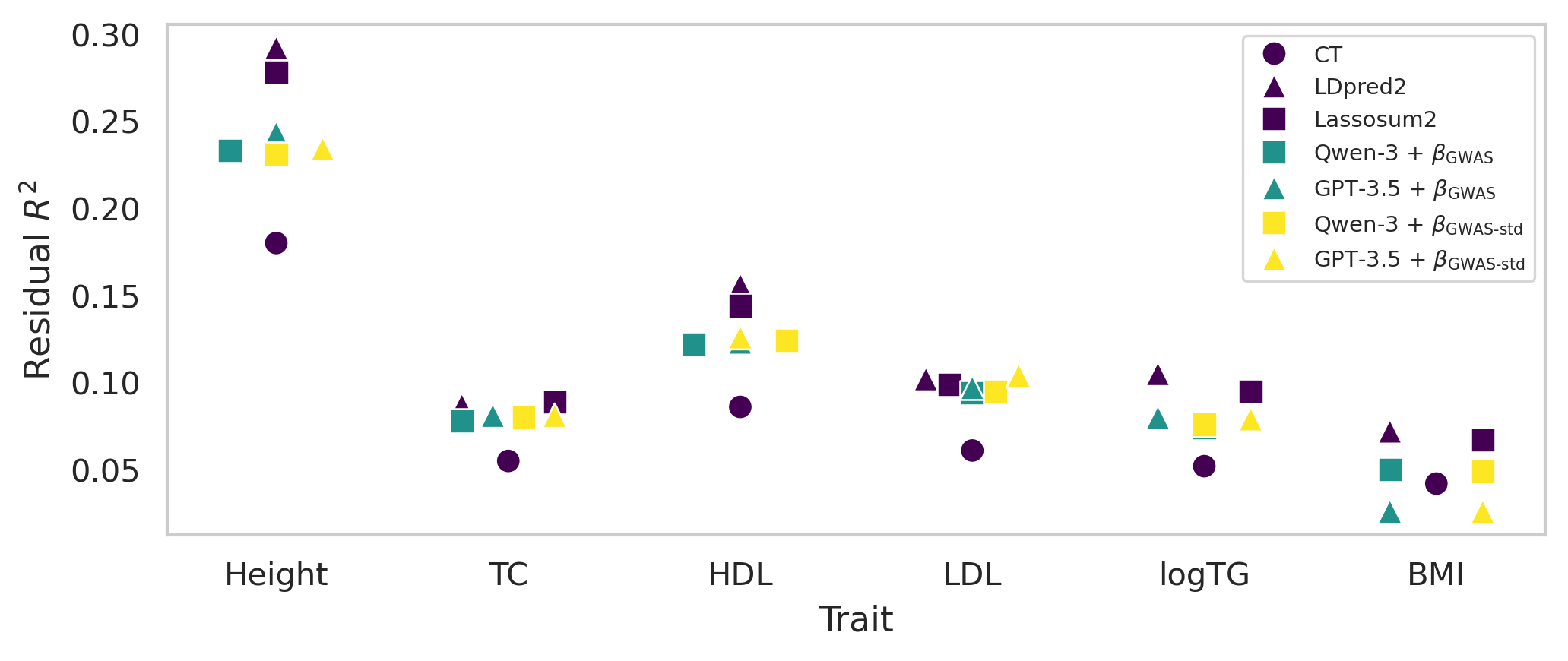}
            \caption{Validation residual $R^2$ adjusted for covariates for six continuous traits predicted from UK Biobank cohort. Traditional PRS pipelines (purple) are compared against LLM-embedding derived predictions using both standardized (yellow) and raw (green) genotype dosages.}
            \label{fig:PRS-embeddings-R2}
            \end{figure}

    In Table \ref{tab:UKB-pred-continous-r2}, we observe that our method using the OpenAI embeddings records the highest test set residual $R^2$ adjusting for covariates for the LDL phenotype. We also note that the predictive performance for the LLM-derived embedding models fit comparably between the PRS methods as seen in \Cref{fig:PRS-embeddings-R2}, where both sets of predictions from the Qwen and OpenAI embeddings model outperform CT in all cases except BMI. Naturally, BMI is a particularly challenging trait as it is highly polygenic. Nevertheless, we find that the GPT-3.5 based prediction slightly outperforms CT. 
   
    \subsubsection{UK Biobank Binary Traits}\label{sec:PRS-UKB-binary}

    For the binary traits: asthma, breast cancer, CAD, prostate cancer, and T2D, the covariates are again age, $\text{age}^2$, sex, and 10 ancestry PCs as is typical in GWAS and PRS studies, except that sex is excluded for modeling prostate cancer and breast cancer. For a brief review of the CT pipeline, we start with the same setup as in the continuous case, defining the PRS for each individual $i$ in the same manner:
        \begin{align*}
            \mathrm{PRS}_i^{T_k} = \sum_{j\in S(T_k)} \widehat{\beta}_j \mathrm{G}_{ij}
            \end{align*}
    where $\mathrm{G}_{ij}$ is the genotype dosage of individual $i$ at SNV $j$, and $\widehat{\beta}_j$ the exact GWAS-derived effect size for variant $j$. Thresholds $T_k$ are then tuned using the PRS scores directly, where confounding is controlled through the evaluation metric by using covariate-adjusted AUC via the RISCA package in R, on the same training, tuning, and validation split. Once the optimal threshold $T^*$ is determined, a final logistic model is fit to determine the estimated effects: 
        \begin{align*}
                \mathrm{logit} \{\mathrm{P}(Y_i=1)\} & = \gamma_0 + \gamma_k\mathrm{PRS}_i^{T^*} +X_i^\top \alpha \\ 
                & = \gamma_0 + \gamma_k\mathrm{PRS}_i^{T^*} + (\mathrm{age}_i, \mathrm{age}^2_i, \mathrm{sex}_i, \mathrm{PC}_{1,i}, \ldots \mathrm{PC}_{10,i})^\top \alpha
            \end{align*}
    where $Y_i$ is the now binary outcome, $\gamma_k$ are the regression coefficients for the PRS and $\alpha$ are the regression coefficients for covariates $X_i$ at the optimal threshold $T^*$. 
    
    To parallel this pipeline, the LLM-aided predictions leverage logistic regression with ridge penalty treating solely the LLM-derived embeddings as features. The same set of potential confounders is now adjusted for in the performance metric through the covariate-adjusted AUC evaluation metric, same as in the CT pipeline, again on the same training/tuning/validation splits. More specifically, we first derive scores $S_i^{T_k}$ from ridge penalized regressions using embeddings derived at different $p$-value thresholds from previous GWAS summary statistics, fitting: 
        \begin{align*}
                S_i^{T_k}=\mathrm{logit} \{\mathrm{P}(Y_i=1)\} & = \widehat{\theta}_0 + \widehat{\theta}_k^\top\mathrm{L}_i^{T_k}
            \end{align*}
    where $S_i^{T_k}$ are the predicted scores at threshold $T_k$, $Y_i$ are the binary outcomes for individual $i$, $\theta_k$ are the regression coefficients, and $\mathrm{L}_i^{T_k}$ are the LLM-derived embeddings at threshold $T_k$ for individual $i$. The optimal threshold $T^*$ is then tuned over the tuning split using covariate-adjusted AUC to control for the role of the confounders directly in the evaluation metric as in the CT pipeline, where the scores $S_i^{T_k}$ take the role of the previous $\mathrm{PRS}_i^{T_k}$. Finally, after $T^*$ is determined, the previous training and tuning splits are combined to fit a final logistic model with ridge penalty over the embeddings, which is used to evaluate over the validation set using covariate-adjusted AUC. Again due to the low sample size of other ancestry groups in the UKB cohort, we report evaluation metrics over individuals of European ancestry in the held-out validation set.

\begin{figure}[ht] 
            \centering
            \includegraphics[width=0.80\linewidth]{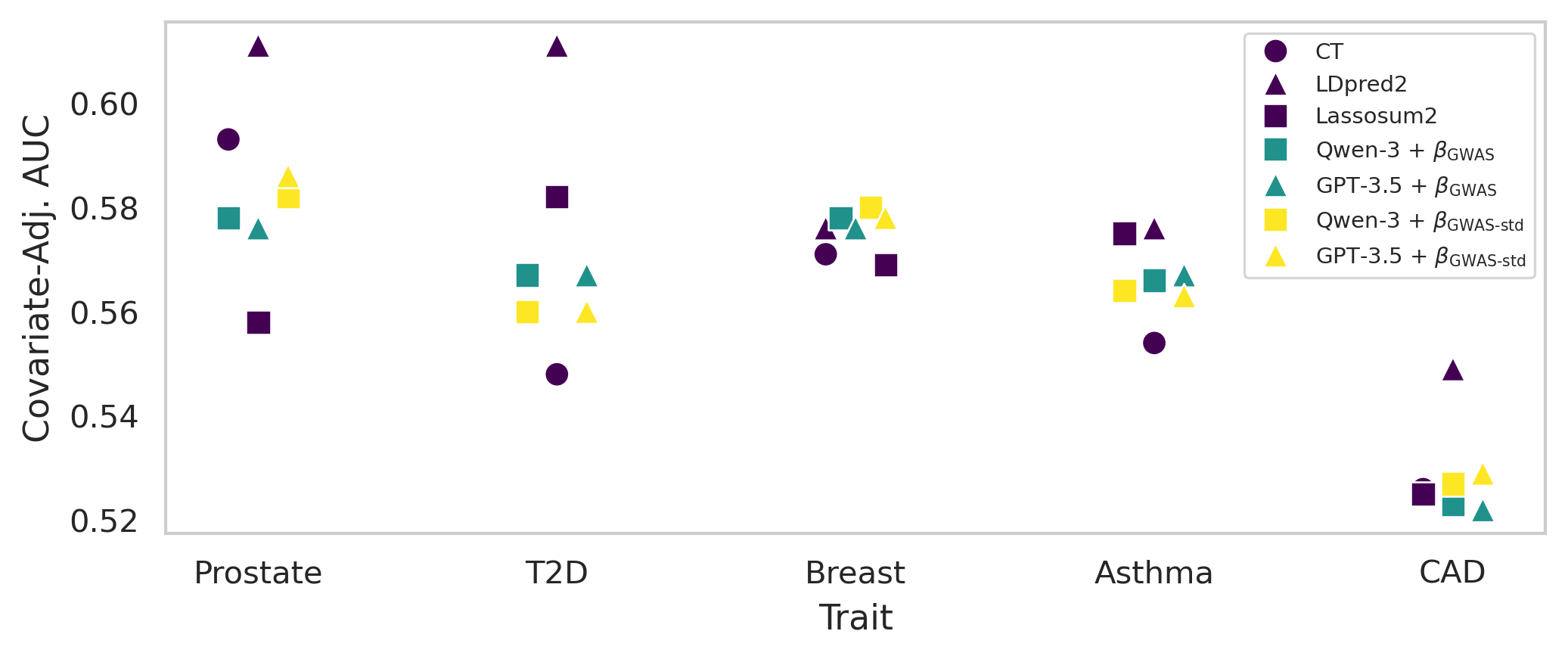}
            \caption{Validation covariate-adjusted AUC for five binary traits predicted from UK Biobank cohort. Traditional PRS pipelines (purple) are compared against LLM-embedding derived predictions using both standardized (yellow) and raw (green) genotype dosages.}
            \label{fig:PRS-embeddings-AUC}
            \end{figure}

    \begin{table}[ht]
        \centering
        \label{tab:r2_methods}
        \begin{tabular}{l c c c c c  }
            \toprule
            Method              & Asthma    & Breast    & CAD      & Prostate  & T2D    \\
            \midrule
            CT                  & 0.554     & 0.571     & 0.526    & 0.593     & 0.548  \\
            LDpred2             & 0.576     & 0.576     & 0.549    & 0.611     & 0.611  \\
            Lassosum2           & 0.575     & 0.569     & 0.525    & 0.558     & 0.582  \\
            \hline
             Qwen-3 + $\beta_{\text{GWAS}}$       & 0.566     & 0.578     & 0.523    & 0.578     & 0.567\\
            GPT-3.5 + $\beta_{\text{GWAS}} $      & 0.567     & 0.576     & 0.522    & 0.576     & 0.567       \\
            Qwen-3 + $\beta_{\text{GWAS-std}}$    & 0.564     & 0.580     & 0.527    & 0.582     & 0.560  \\
            GPT-3.5 + $\beta_{\text{GWAS-std}} $  & 0.563     & 0.578     & 0.529    & 0.586     & 0.560       \\
            \bottomrule 
        \end{tabular}
        \caption{Predictive performance (covariate-adjusted AUC) of different PRS methods in the held-out test set  over individuals of European ancestry compared to LLM-derived embeddings based methods. LLM-derived embeddings include both those using non-standardized and standardized genotype dosages for comparison. The specific models used are \texttt{Qwen3-Embedding-0.6B} and \texttt{text-embedding-3-large}.}
        
        \label{tab:UKB-pred-binary-auc}
    \end{table}
    In \Cref{tab:UKB-pred-binary-auc} and \Cref{fig:PRS-embeddings-AUC}, we observe that our method using the both the OpenAI and Qwen3 embeddings with GWAS effect weighting performs comparably within the range of the three traditional PRS methods for all outcomes, adjusting for both individual covariates and ancestry PCs. Furthermore, our GWAS-weighted LLM-aided method further outperforms all the PRS methods on the breast cancer trait using both standardized and non-standardized genotype dosages. 

\subsection{Dimension Reduction of LLM Embeddings}
    As in \Cref{sec:GWAS-ancestry-adjustment}, dimension reduction is often applied through PCA to genotype matrices for tasks such as ancestry adjustment in GWAS. Hence, it is crucial to better understand the role of dimension in the case of these LLM-derived individual-level embeddings. Here, we explore the role of the top 100 principal components as it relates to the cumulative variance explained within each set of embeddings, as well as performance on the downstream prediction tasks of the previous section. 
    
\subsubsection{UKB Continuous Traits}

\begin{figure}[ht!]
            \centering
            \begin{subfigure}{0.40\linewidth}
                \centering
                \includegraphics[width=\linewidth]{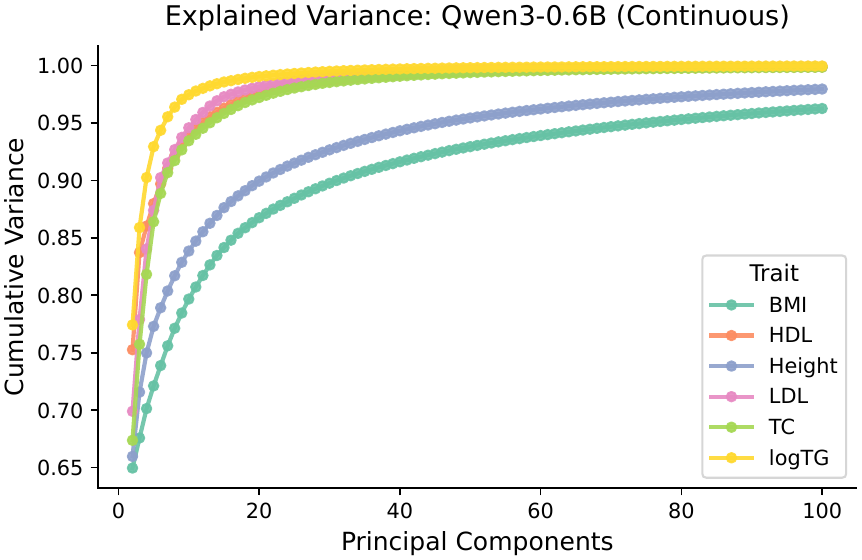}
                \caption{Qwen3-Embedding-0.6B}
                \label{fig:PCA-emb-qwen3-a}
                \end{subfigure}
            \begin{subfigure}{0.40\linewidth}
                \centering
                \includegraphics[width=\linewidth]{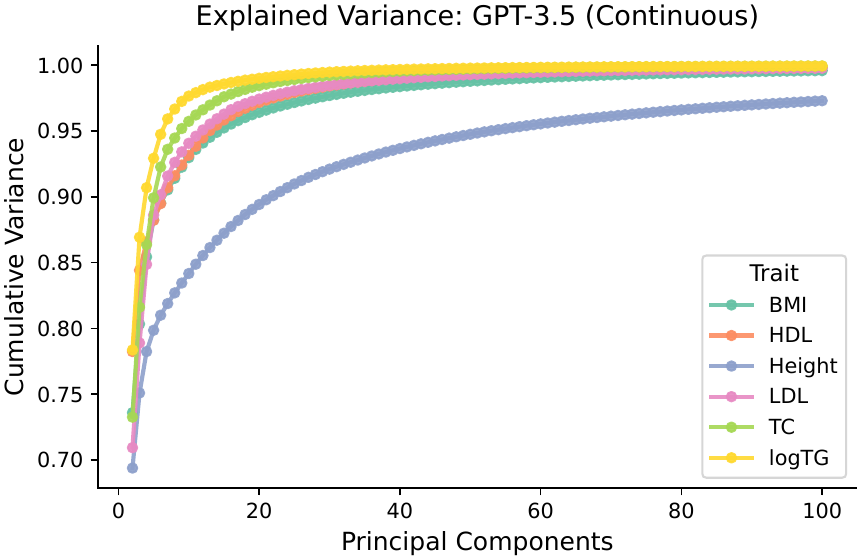}
                \caption{Text-Embedding-3-Large}
                \label{fig:PCA-emb-gpt3-b}
                \end{subfigure}
            \caption{Top 100 principal component cumulative variance plot for a) \texttt{Qwen3-Embedding-0.6B} and b) OpenAI \texttt{text-embedding-3-large} across embeddings for 6 continuous traits at the previously optimized $p$-value threshold for the prediction task. Here, individual-embeddings are weighted by trait-specific GWAS effect sizes using standardized genotype dosages.}
            \label{fig:PCA-explained-both-continuous}
            \end{figure}

  Here we present the results on assessing the top 100 principal components over the embeddings used in the continuous trait UK Biobank prediction task. Embeddings are taken at the optimized thresholds in \Cref{sec:PRS-UKB-continuous} using the standardized genotype dosage weights. In \Cref{fig:PCA-explained-both-continuous} we find that the top 100 PCs explains the vast majority of variance across all six continuous traits, whereas about top 10 PCs are sufficient to explain the vast majority of variance for lipid traits (HDL, LDL, TC and logTG).
    
     \begin{figure}[ht!]
        \centering
        \begin{subfigure}{0.3\textwidth}
            \includegraphics[width=\textwidth]{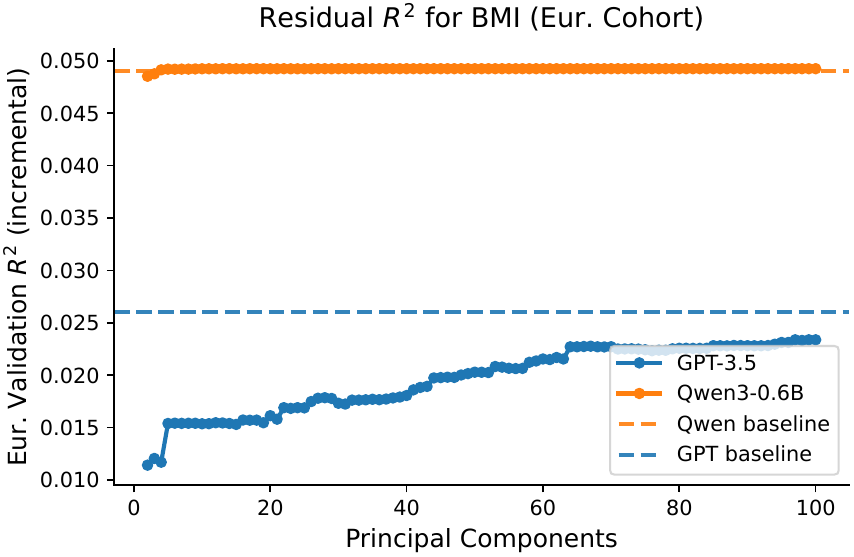}
        \end{subfigure}%
        \begin{subfigure}{0.3\textwidth}
            \includegraphics[width=\textwidth]{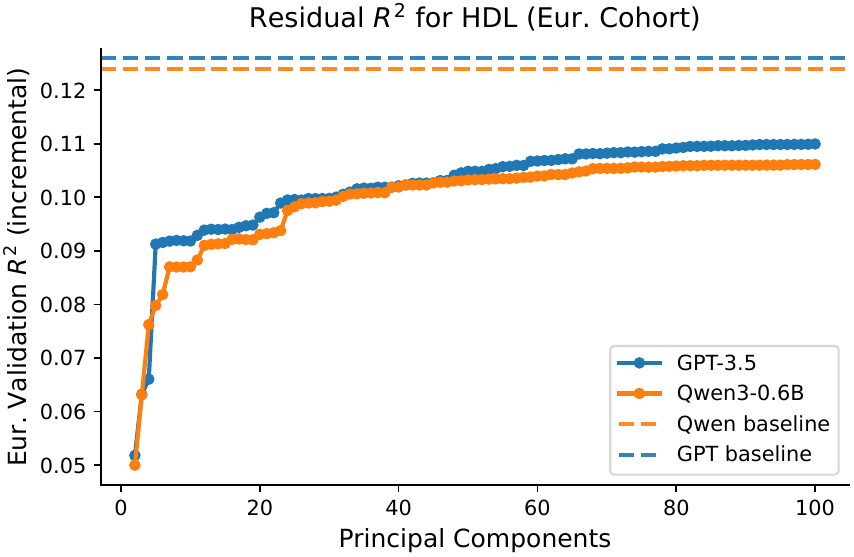}
        \end{subfigure}%
        \begin{subfigure}{0.3\textwidth}
            \includegraphics[width=\textwidth]{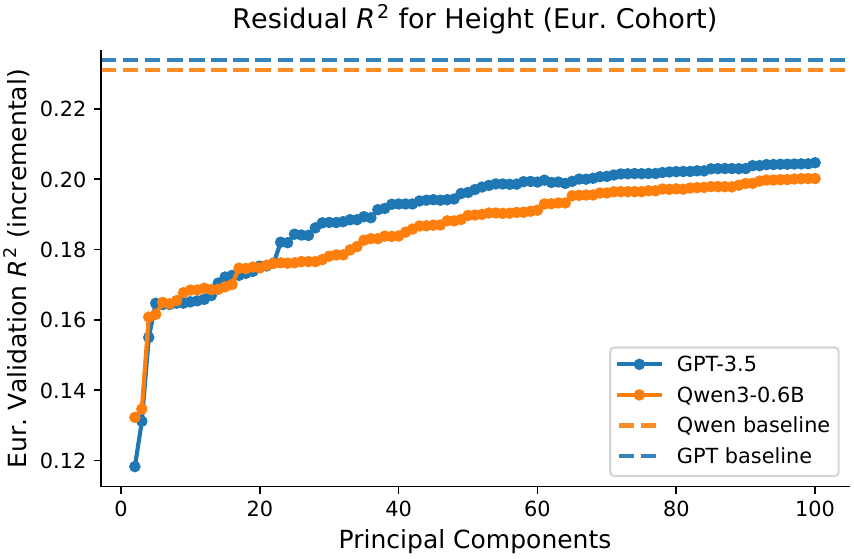}
        \end{subfigure}\\
        \begin{subfigure}{0.3\textwidth}
            \includegraphics[width=\textwidth]{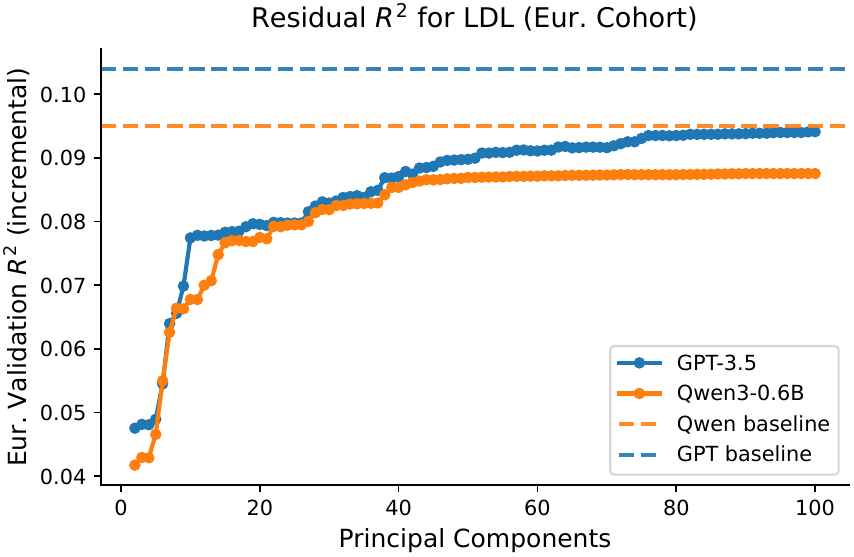}
        \end{subfigure}%
        \begin{subfigure}{0.3\textwidth}
            \includegraphics[width=\textwidth]{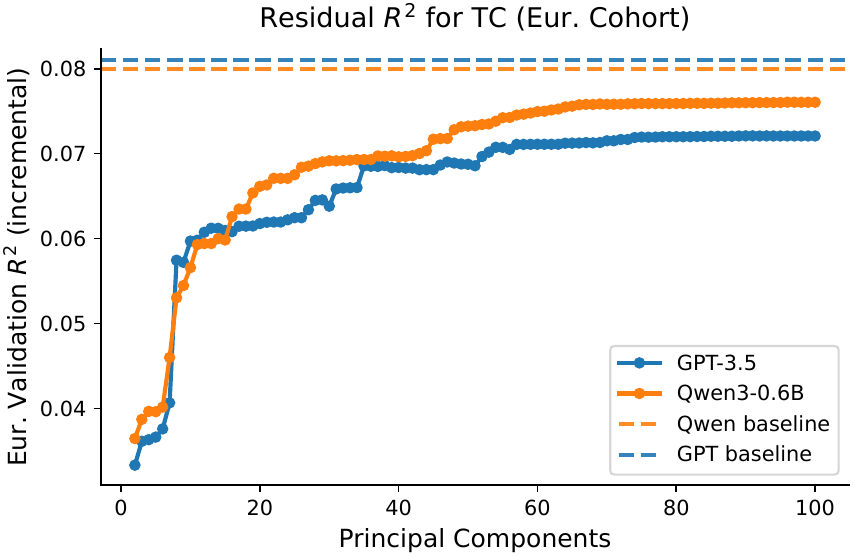}
        \end{subfigure}
        \begin{subfigure}{0.3\textwidth}
            \includegraphics[width=\textwidth]{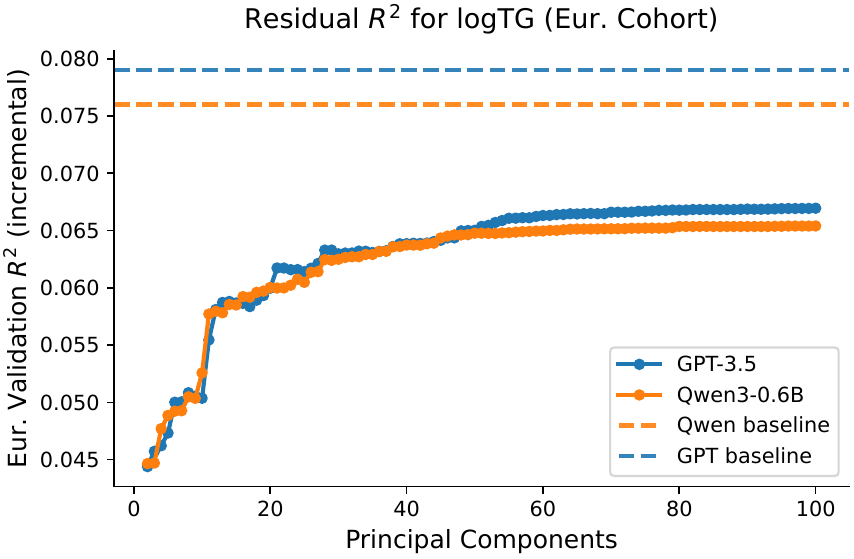}
        \end{subfigure}
    \caption{Residual $R^2$ after confounder adjustment evaluated on fitting ridge regression models to regression tasks in UK Biobank cohort using LLM-derived embeddings where the dimension is first further reduced using PCA to the top 2-100 principal components. Embeddings used are based on the optimized thresholds from the PRS-style analysis with the standardized genotype dosage weighting. Performance is evaluated on a held-out test set separate from model fitting.}
    \label{fig:PCA-emb-continuous-traits}
    \end{figure}

    In \Cref{fig:PCA-emb-continuous-traits}, we find that for the continuous traits, performance on the incremental $R^2$ adjusting for baseline covariates and ancestry PCs has a generally monotonically increasing relationship with the number of principal components used. For all traits, we see significantly improved performance when incorporating between 10-20 PCs. Finally, we observe that using the top 100 PCs (compared to the full dimension of 1024 or 3072 for the LLM embeddings) is not always enough to attain the performance of using the full model such as in height and logTG in particular.

\subsubsection{UKB Binary Traits}
\begin{figure}[ht!]
            \centering
            \begin{subfigure}{0.40\linewidth}
                \centering
                \includegraphics[width=\linewidth]{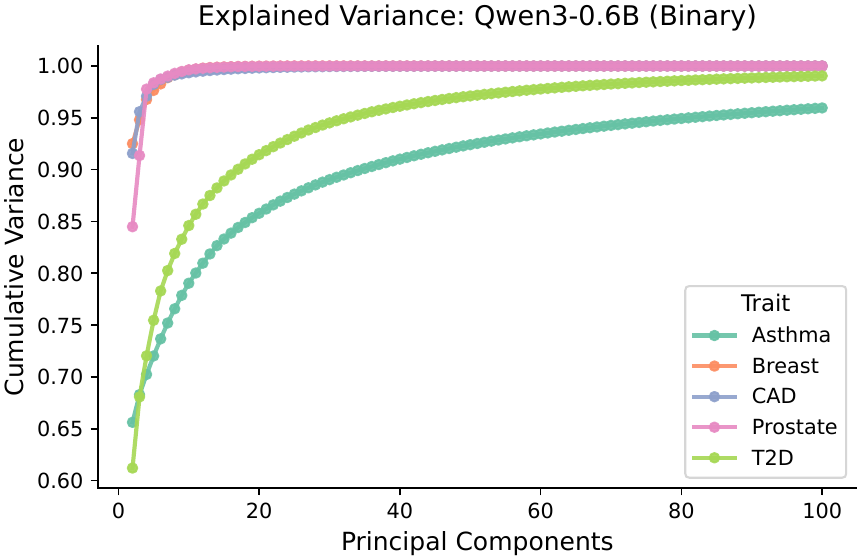}
                \caption{Qwen3-Embedding-0.6B}
                \label{fig:PCA-emb-qwen3-a}
                \end{subfigure}
            \begin{subfigure}{0.40\linewidth}
                \centering
                \includegraphics[width=\linewidth]{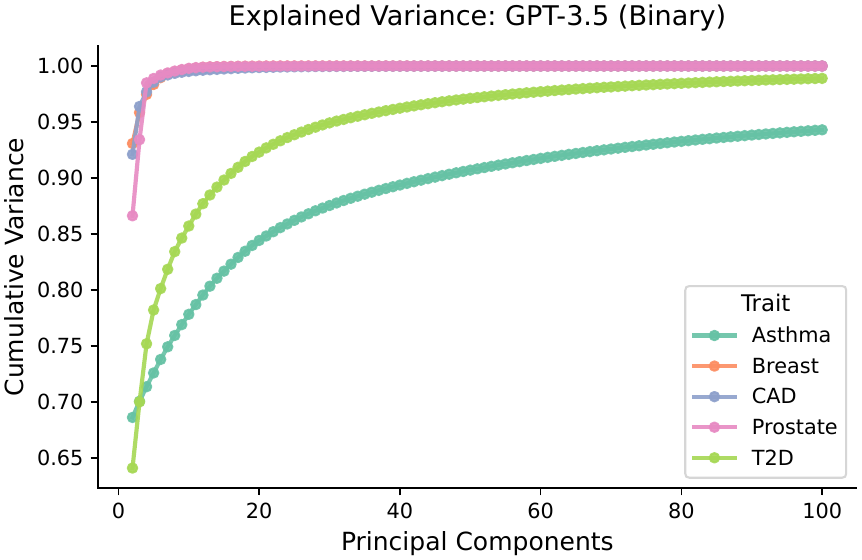}
                \caption{Text-Embedding-3-Large}
                \label{fig:PCA-emb-gpt3-b}
                \end{subfigure}
            \caption{Top 100 principal component cumulative variance plot for a) \texttt{Qwen3-Embedding-0.6B} and b) OpenAI \texttt{text-embedding-3-large} across embeddings for 5 binary traits at the previously optimized $p$-value threshold for the prediction task. Here, individual-embeddings are weighted by trait-specific GWAS effect sizes using standardized genotype dosages.}
            \label{fig:PCA-explained-both-binary}
            \end{figure}

    Here we present the results on assessing the top 100 principal components over the embeddings used in the binary trait UK Biobank prediction task. Embeddings are taken at the optimized thresholds in \Cref{sec:PRS-UKB-binary} using the standardized genotype dosage weights. In \Cref{fig:PCA-explained-both-binary} we find that the top 100 PCs explains the vast majority of variance across all 5 binary traits, whereas about top 10 PCs are sufficient to explain the vast majority of variance for the CAD, prostate, and breast cancer traits. 
    
    \begin{figure}[ht!]
        \centering
        \begin{subfigure}{0.3\textwidth}
            \includegraphics[width=\textwidth]{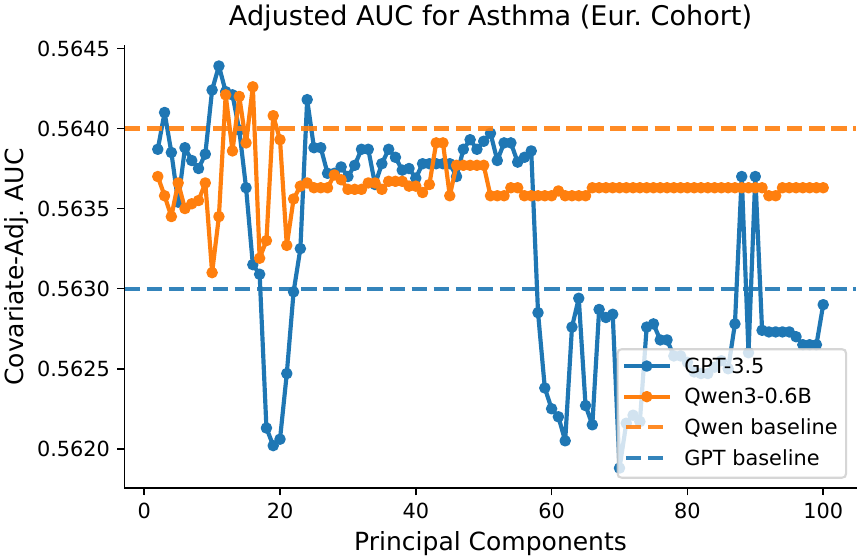}
        \end{subfigure}%
        \begin{subfigure}{0.3\textwidth}
            \includegraphics[width=\textwidth]{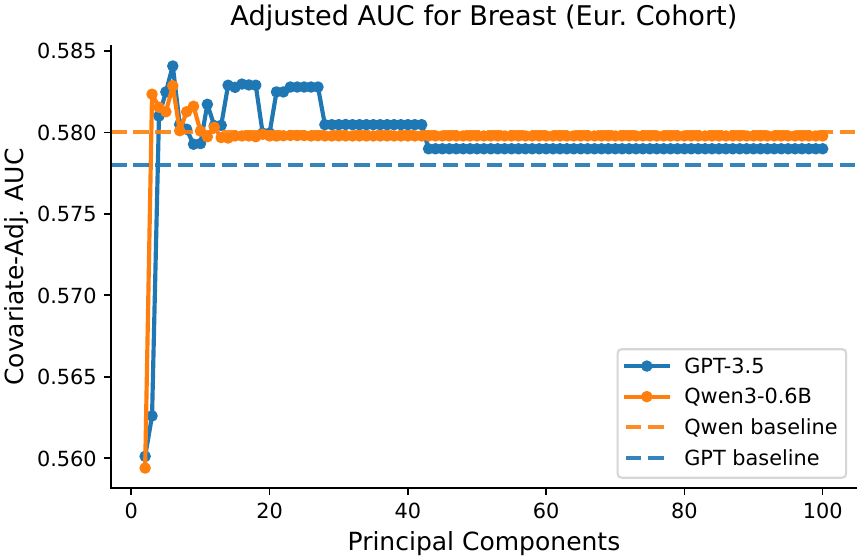}
        \end{subfigure}%
        \begin{subfigure}{0.3\textwidth} 
            \includegraphics[width=\textwidth]{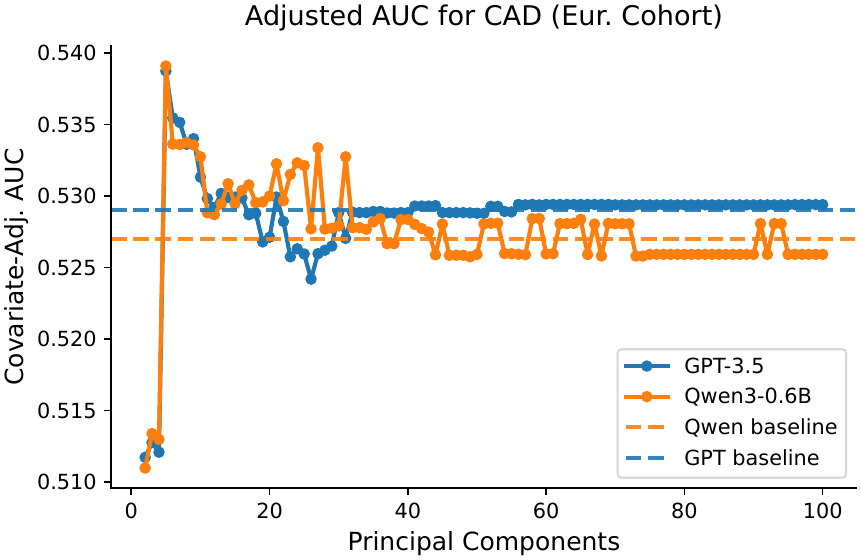}
        \end{subfigure}\\
        \begin{subfigure}{0.3\textwidth}
            \includegraphics[width=\textwidth]{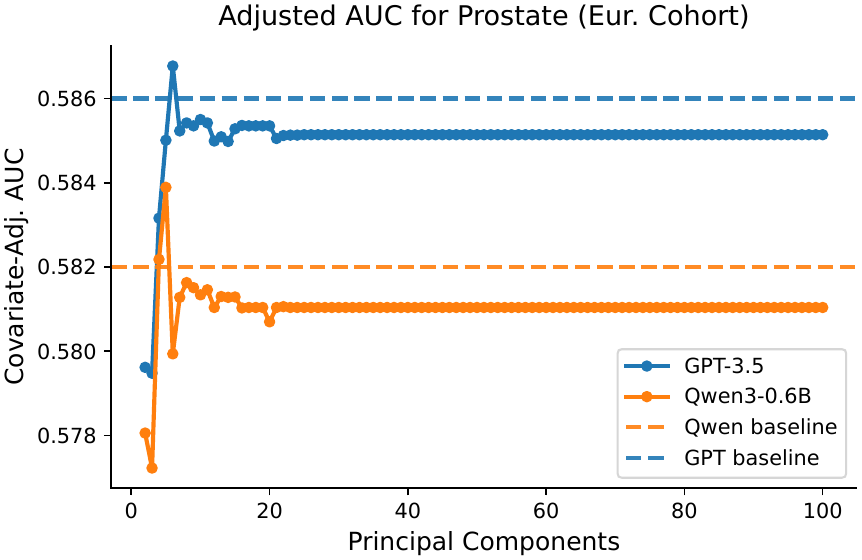}
        \end{subfigure}%
        \begin{subfigure}{0.3\textwidth}
            \includegraphics[width=\textwidth]{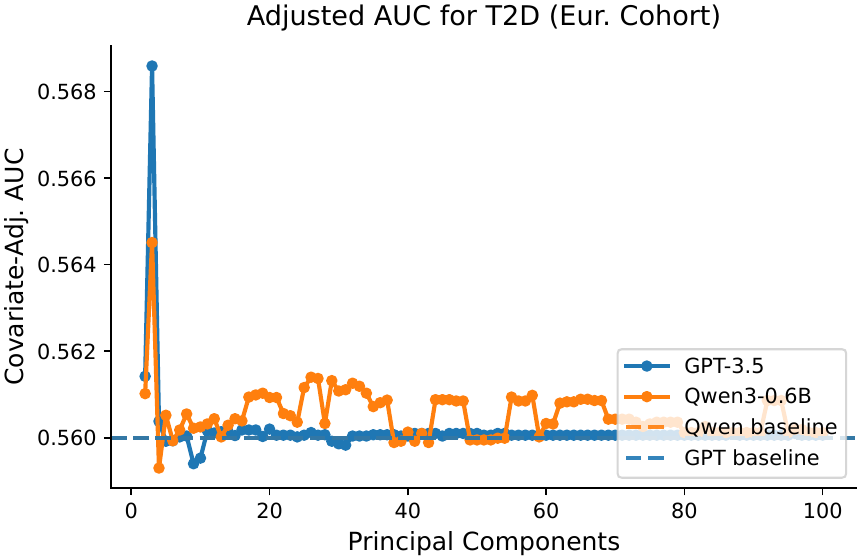}
        \end{subfigure}
    \caption{Covariate-adjusted AUC evaluated on fitting ridge regression models to binary classification tasks in UK Biobank cohort using LLM-derived embeddings where the dimension is first further reduced using PCA to the top 2-100 principal components. Embeddings are used based on the optimized thresholds from the PRS-style analysis with the standardized genotype dosage weighting. Performance is evaluated on a held-out test set separate from model fitting.}
    \label{fig:PCA-emb-binary-traits}
    \end{figure}
            
    Next, in \Cref{fig:PCA-emb-binary-traits}, we observe that generally for very low principal component dimensions, the performance on the prediction tasks at a given GWAS-derived $p$-value cutoff can fluctuate, however, performance across all traits tends to plateau after adding around 20-30 principal components. 
    
    Although the predictive performance tends to increase with PC dimension, particularly for continuous traits, performance did not reach that of the full model for most traits when using the first 100 PCs with noticeable differences such as height and logTG. For binary traits, we observe a relationship where using fewer PCs can offer better performance than the comparative full model, suggesting that the best balance between model dimension, predictive performance, and computational efficiency is highly trait specific. Together, this corroborates the need for flexible approaches to dimension reduction when working with genetic data, highlighting the computational advantage of our embeddings approach where the full dimension is fixed to the architecture of the LLM embeddings model. Compared to raw biobank data where the dimension (i.e. number of genetic variants) can vary greatly and is often hundreds of times larger, the use of LLM embeddings with fixed dimension offers practical and significant speedups for computing PCs.

\section{Discussion and Future Work}

    The study of population-scale genetics with quantitative and disease traits has significant public health relevance, offering a way to assess an individual's genetic risk for a variety of potentially significant health outcomes such as diabetes and cancer, without the need for invasive lab testing. However, because genetic effects are often distributed over many variants across the genome, it can frequently involve a challenging, high-dimensional prediction task. Current PRS pipelines therefore must balance how many variants to use, while remaining computationally feasible. Furthermore, traditional PRS pipelines primarily focus on using genetic variants with evidence from statistical associations, and do not necessarily take advantage of the background information available on many of the well-studied variants for which there is a rich literature. 

    To address this, we have curated functional annotations at 3 different resolutions covering all possible 8.9 billion single nucleotide polymorphisms in the human genome, using high-quality data from the FAVOR \citep{zhou2023favor, zhou2026favor}, ClinVar \citep{landrum2016clinvar}, and GWAS Catalog \citep{buniello2019nhgri, cerezo2025nhgri} datasets. From these functional annotations consisting of information from in silico prediction models, clinically verified traits, as well as statistical associations from peer-reviewed publications, we derived general purpose LLM embeddings using OpenAI’s \texttt{text-embedding-3-large} and Qwen's \texttt{Qwen3-Embedding-0.6B} capturing the semantic information available from our embeddings while also leveraging extensive pre-training accessible to these large foundational models. Notably, these are among the first pre-computed embeddings at the genetics variant level, saved, and released as datasets for public use. 
    
    We release all three datasets described, namely the 1.5 million SNV embeddings dataset from HapMap3~\citep{international2010integrating} $+$ Multi-Ethnic Genotyping Arrays (MEGA,~\citealp{bien2016strategies}) using OpenAI's \texttt{text-embedding-3-large}, UK Biobank Imputed 90 million variant list also using using OpenAI's \texttt{text-embedding-3-large}, as well as the full set of 8.9 billion variants available from FAVOR~\citep{zhou2023favor} using \texttt{Qwen3-Embedding-0.6B}, with the link to the public Hugging Face repository available in the Appendix. 
    
    Along with the preliminary prediction tasks serving as basic quality checks, we also demonstrate the effectiveness of the LLM-derived embeddings in PRS-style tasks using solely linear or generalized linear modeling with ridge penalty over the individual-level embeddings, which performs on par or even better than modern PRS pipelines including CT \citep{wray2007prediction, international2009commonCT}, LDpred2 \citep{prive2020ldpred2} and Lassosum2 \citep{prive2022lassosum2} on a range of continuous and binary traits from the UK Biobank cohort. Due to the modularity of the pipeline, stronger machine learning-based prediction models may see even larger improvements. 

    We plan a few lines of future work to address the potential challenges and other applications of using these LLM-derived embeddings. First, the second of edition of FAVOR has since been released \citep{zhou2026favor}, offering a even richer source of high quality to generate variant-level annotations and embeddings. Second, our current method of obtaining individual-level embeddings is to weight the variant embeddings by genotype dosage (i.e. 0, 1, or 2) or by incorporating the GWAS effect estimates. Further work can be done to assess different ways of more meaningfully capturing relationships between variants such as using LD. Third, conventional PRS approaches aggregate effect-size-weighted allele counts across common genetic variants~\citep{wray2007prediction,international2009commonCT,prive2020ldpred2,ge2019polygenic,mak2017polygenic}. A key feature is that the resulting weights can be interpreted as effect size estimates for each variant. Currently under our pipeline, the individual-level embeddings are no longer as interpretable as traditional PRS. However, now that we have shown the embeddings can themselves be useful in risk prediction, we plan to incorporate them more directly to augment traditional PRS pipelines to further improve prediction while retaining key interpretability. Lastly, given our variant-level embeddings are precomputed for all possible variants, we will also expand our PRS-style prediction pipeline by integrating both common and rare variants following workflows such as RICE \citep{williams2024integrating}.

    In summary, we have curated one of the first large-scale, LLM-based embeddings datasets over functional annotations of all possible 8.9 billion genetic variants in the human genotype. These embeddings allow for highly flexible application to downstream tasks such as offering a computationally efficient way of performing ancestry adjustment in GWAS pipelines with highly concordant results compared to traditional methods, and competitive performance when used for PRS-style prediction tasks compared to state-of-the-art methods in biobank-scale analysis. Finally, these pre-computed variant embeddings are fully released publicly at three different resolutions including the full 8.9 billion set with the direct link in the Appendix.

\subsection*{Acknowledgments}
HN was supported by the National Institutes of Health (NIH) R01 CA253450; XL is supported by NIH R01 HL173044, R01 AG085581. DL was supported by NIH R01 LM014407, P50 CA058223, P42 ES031007. DL acknowledges support from the NVIDIA Academic Grant Program and support from OpenAI API credits provided through the National Artificial Intelligence Research Resource (NAIRR) Pilot. DL and XL acknowledge a Pivot Award from the University of North Carolina at Chapel Hill Gillings Research Office.
    
The analysis utilized the high-performance computation Biowulf cluster at NIH, USA. This research has been conducted using the UK Biobank Resource under Application Number 52008. We gratefully acknowledge UK Biobank participants for their contributions, without whom this research would not have been possible. This research was supported in part by the Intramural Research Program of the NIH. The contributions of the NIH authors were made as part of their official duties as NIH federal employees, are in compliance with agency policy requirements, and are considered Works of the United States Government. However, the findings and conclusions presented in this paper are those of the authors and do not necessarily reflect the views of the NIH or the U.S. Department of Health and Human Services.

\appendix
\section*{Appendix}

\section{Data Availability}
    All variant-level embeddings used in the study and described above are currently released on HuggingFace at the aforementioned 3 different resolutions: 
    \begin{itemize}
        \item \url{https://huggingface.co/datasets/LiLabUNC/Variant-Foundation-Embeddings}. 
        \end{itemize}

\section{Computational Costs}

    Due to the size and scale of the genetic variant embeddings, there are significant computational costs associated with generating the embeddings. For the medium-sized $\sim$90 million UKB Imputed variant list, we estimate a cost of approximately $\$1,000$ using the OpenAI batched API, while the GPU hours required for using even the smallest Qwen3-Embedding model over the full $\sim$9 billion variant list required approximately $5,000$ hours running on Nvidia L40 GPUs. In Table \ref{tab:asdf} and Table \ref{tab:asdf2}, we present estimated costs and storage for all datasets. 

        \begin{table}[ht!]
            \centering
            \small
            \begin{tabular}{l r r r}
                \toprule
                Dataset (GPT3) & Variant List & Size & API Cost  \\
                \midrule
                 HapMap3/MEGA & 1.5 million & 17.5 GB & \$15\\
                 UKB Imputed & 90 million & $\sim$1 TB& $\sim$\$1,000\\
                 Full & 9 billion & $\sim$100TB & $\sim$\$100,000\\
                \bottomrule
                \end{tabular}
            \caption{Estimated costs from using the OpenAI batched API for the \texttt{text-embedding-3-large} embeddings model.}
            \label{tab:asdf}
            \end{table}

        \begin{table}[ht]
            \centering
            \small
            \begin{tabular}{l r r r}
                \toprule
                Dataset (Qwen3) & Variant List & Size & GPU Hours  \\
                \midrule
                 HapMap3/MEGA & 1.5 million & 3.7 GB & 3\\
                 UKB Imputed & 90 million & $\sim$250 GB & $\sim$200\\
                 Full & 9 billion & $\sim$30 TB & $\sim$5,000\\
                \bottomrule
                \end{tabular}
            \caption{Estimated costs for generating open-source embeddings using \texttt{Qwen3-Embedding-0.6B} model with Nvidia L40 GPUs.}
            \label{tab:asdf2}
            \end{table}

\bibliographystyle{chicago}
\bibliography{ref}

\end{document}